\documentclass[twocolumn,superscriptaddress,aps,prl,showpacs,preprintnumbers,amsmath,amssymb,floatfix]{revtex4-1}

\usepackage[latin9]{inputenc}
\usepackage{color}
\usepackage{dcolumn}
\usepackage{bm}
\usepackage{bbm}
\usepackage{braket}
\usepackage{mathrsfs}
\usepackage{amstext}

\usepackage{txfonts}

\usepackage{amsmath}
\usepackage{dsfont}
\usepackage{euscript}
\usepackage{amssymb}
\usepackage{graphicx}
\usepackage{esint}
\usepackage[unicode=true,
 bookmarks=false,
 breaklinks=false,pdfborder={0 0 1},backref=false,colorlinks=true,citecolor=blue]
 {hyperref}

\makeatletter
\@ifundefined{textcolor}{}
{%
 \definecolor{BLACK}{gray}{0}
 \definecolor{WHITE}{gray}{1}
 \definecolor{RED}{rgb}{1,0,0}
 \definecolor{GREEN}{rgb}{0,1,0}
 \definecolor{BLUE}{rgb}{0,0,1}
 \definecolor{CYAN}{cmyk}{1,0,0,0}
 \definecolor{MAGENTA}{cmyk}{0,1,0,0}
 \definecolor{YELLOW}{cmyk}{0,0,1,0}
}

\newcommand{\eref}[1]{Eq.\,\eqref{#1}}
\newcommand{\fref}[1]{Fig.\,\ref{#1}}

\makeatother

\begin{document}

\title{Entanglement and spin-squeezing without infinite-range interactions}

\author{Michael Foss-Feig}
\affiliation{United States Army Research Laboratory, Adelphi, MD 20783, USA}
\affiliation{Joint Quantum Institute, NIST/University of Maryland, College Park, MD 20742 USA}
\affiliation{Joint Center for Quantum Information and Computer Science, NIST/University of Maryland, College Park, MD 20742 USA}

\author{Zhe-Xuan Gong}
\affiliation{Joint Quantum Institute, NIST/University of Maryland, College Park, MD 20742 USA}
\affiliation{Joint Center for Quantum Information and Computer Science, NIST/University of Maryland, College Park, MD 20742 USA}

\author{Alexey V. Gorshkov}
\affiliation{Joint Quantum Institute, NIST/University of Maryland, College Park, MD 20742 USA}
\affiliation{Joint Center for Quantum Information and Computer Science, NIST/University of Maryland, College Park, MD 20742 USA}

\author{Charles W. Clark}
\affiliation{Joint Quantum Institute, NIST/University of Maryland, College Park, MD 20742 USA}

\begin{abstract}
Infinite-range interactions are known to facilitate the production of highly
entangled states with applications in quantum information and
metrology.  However, many experimental systems have interactions that decay with distance, and the achievable benefits in this context are much less clear. Combining recent exact solutions with a controlled
expansion in the system size, we analyze quench dynamics in
Ising models with power-law ($1/r^{\alpha}$) interactions in $D$
dimensions, thereby expanding the understanding of spin
squeezing into a broad and experimentally relevant context.  In
spatially homogeneous systems, we show that for small
$\alpha$ the scaling of squeezing with system size is \emph{identical} to the infinite-range
($\alpha=0$) case.  This indifference to the interaction range
persists up to a critical value $\alpha=2D/3$, above which
squeezing degrades continuously.  Boundary-induced inhomogeneities
present in most experimental systems modify this picture, but it nevertheless remains qualitatively correct for finite-sized systems.
\end{abstract}


\maketitle


The dynamical growth of entanglement in far-from-equilibrium quantum
systems makes them attractive for numerous applications in quantum
information science, but greatly complicates their description.
For sufficiently short-ranged interactions, entanglement
growth after a quench is constrained to within a light cone \cite{lieb72},
often allowing for efficient numerical simulations
\cite{Calabrese2005,schachenmayer13}.  On the other extreme, systems with infinite-range (i.e.\,distance independent) interactions develop
entanglement on all length scales simultaneously, but are nevertheless
easily characterized in terms of a small set of collective variables \cite{Dicke1954}.
The intermediate territory of long but \emph{not}
infinite-ranged interactions is comparatively uncharted, and
questions surrounding the growth of entanglement---and the associated
computational complexity---in this regime have recently attracted significant theoretical
\cite{hastings05,schachenmayer13,hauke13,Gong14,fossfeig15, eisert13,Eldredge16} and
experimental \cite{yan13,richerme14,lanyon14} attention.

An important application of quench-induced entanglement is the generation of spin-squeezing \cite{ueda}, which has been sought as a means towards
quantum-enhanced metrology
\cite{Wineland1994,esteve,Lucke2011,Chen2011}, and more
generally provides a theoretically tractable and
experimentally accessible metric for
characterizing entanglement in non-equilibrium
systems \cite{fossfeig2013,lee2013,Gil2013,sorensen2001,Sorensen2000,Zhang2013}.
Strategies to produce spin-squeezed states \cite{sorensen2001,PhysRevLett.85.1594,fossfeig2012,PhysRevLett.110.120402,PhysRevLett.116.053601} and the experiments
that have pursued them \cite{esteve,gross,Lucke2011,Monz2011,Chen2011,Bohnet2013} are
extremely diverse, but tend to share the idealization of
infinite-range interactions and perfectly collective dynamics.  Indeed, the difficulty of studying
non-equilibrium systems with long (but not infinite) ranged
interactions has obscured the extent to which squeezing is robust to relaxing this idealization.  This
question is, however, of immediate importance to many long-range
interacting experimental
systems, including polar
molecules \cite{yan13,Hazzard2014}, magnetic atoms \cite{dePaz2013},
Rydberg atoms \cite{RevModPhys.82.2313,nature11596}, trapped ions
\cite{richerme14,lanyon14,Britton12,Bohnet16}, and optical lattice clocks
\cite{Martin2013}, in which interactions decay with distance.

\begin{figure}[!thbp]
\begin{center}
\includegraphics[width = 0.94 \columnwidth]{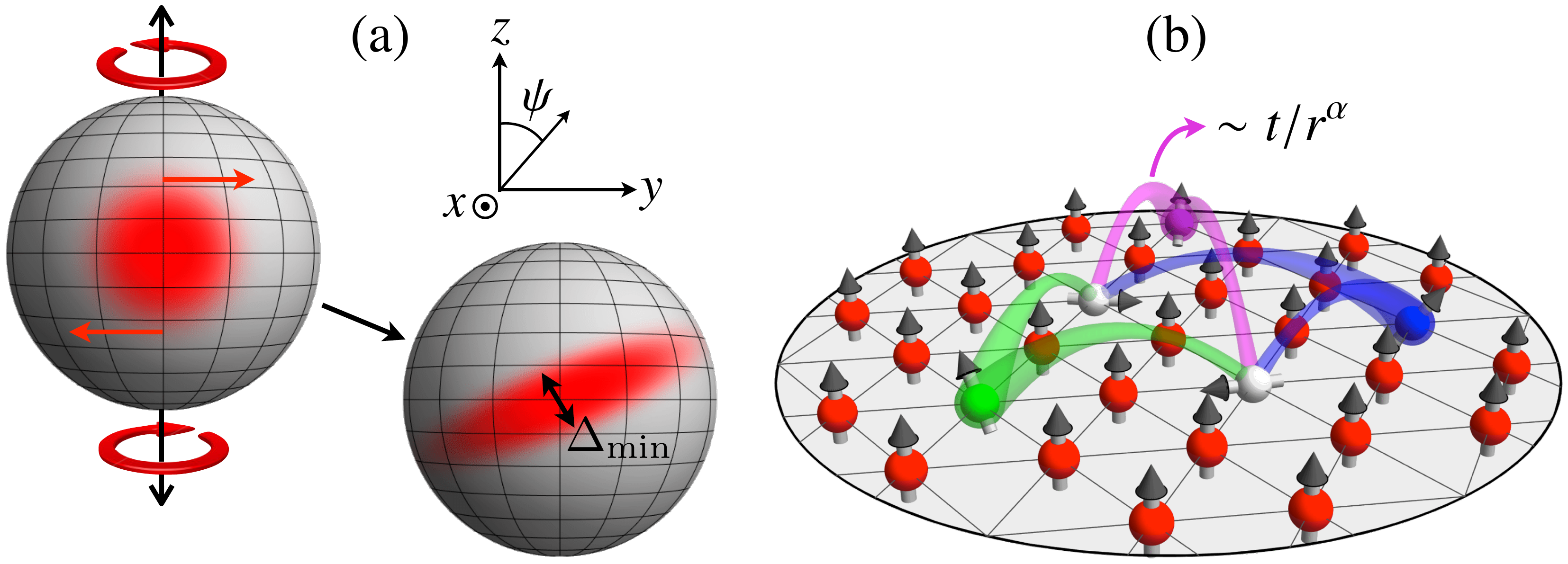}
\caption{(Color online) (a) Spin squeezing in the single-axis twisting
  model depicted on the Bloch sphere. The shaded areas represent the uncertainty due to spin projection noise.  (b) A physical process contributing to the
  correlation between two spins (in white) via three
  intermediate spins. Each leg connecting a white spin to a colored one caries a factor $t/r^{\alpha}$, with $r$ the distance between the spins.}
\label{fig:squeezing_and_correlation_graphic}
\end{center}
\vspace{-1.0 cm}
\end{figure}

To answer this question, we consider a generalization of the \emph{single-axis twisting
model} \cite{ueda} where the spin-spin couplings decay as a
power law $1/r^{\alpha}$, and analytically compute how the amount of squeezing scales with the number of spins $\mathcal{N}$.
Remarkably, we find that for spatially homogeneous systems, the scaling of
squeezing with system size is \emph{completely} unaffected by the
decay of interactions for $\alpha<2D/3$, with $D$ the dimension of space.  In the single-axis twisting
model ($\alpha=0$), quantum correlations between pairs of spins conspire
to reduce spin projection noise [Fig.~\ref{fig:squeezing_and_correlation_graphic}(a)] below the
standard quantum limit by a multiplicative factor $\xi$, called the
squeezing parameter.  For $\alpha>0$, the dynamics are not collective
and the spin-state cannot be represented on a Bloch sphere, but the
indifference of squeezing to interaction range can be understood by
considering the physical processes that generate such correlations.  An example
is shown in Fig.~\ref{fig:squeezing_and_correlation_graphic}(b), where
correlations between two spins (in white) are mediated by three
other spins (colored).  Each leg represents one application of the Hamiltonian and hence caries a
factor $\sim t/r^{\alpha}$, where $t$ is time and $r$ is the separation between the spins
connected by that leg.  For sufficiently small $\alpha$ we can restrict our
attention to processes where the two white spins are minimally
coupled---i.e. by exactly two legs---to each spin mediating their
correlation, since these contributions to the correlation diverge most strongly with
$\mathcal{N}$.  Summing over the position of the mediating spin we find a contribution $\sim t^2\int d^{D}r/r^{2\alpha}\sim
\mathcal{N}\times t^2\mathcal{N}^{-2\alpha/D}$ (the final step here follows from dimensional analysis, and noting that the linear system size scales as $\mathcal{N}^{1/D}$).  Because $t$ always enters in this manner, all correlation functions can be written in terms of $\mathcal{N}$ and a rescaled time $\tau=t\mathcal{N}^{-\alpha/D}$.  Therefore, even though $\xi$ depends on $\mathcal{N}$, $t$ and $\alpha$, we can approximate it as a universal function $\xi(\mathcal{N},\tau)$ of the system size and a rescaled time, with no reference to the range of interactions.  It follows that, at least for sufficiently small $\alpha$, the
decay of interactions only affects the \emph{time} at which the optimal squeezing occurs, but does not affect \emph{how much} squeezing can be achieved.  In what follows, this claim will be substantiated with an explicit calculation, and we will determine the critical value of $\alpha$ above which this simple picture breaks down.  Before proceeding, we note that most experimental systems are not spatially homogeneous (owing to boundary
effects), and for sufficiently large $\mathcal{N}$ such inhomogeneity
will be seen to impose a finite minimal value for $\xi$ that is independent of system size. Nevertheless, we will demonstrate
that the qualitative picture just described persists in relatively large (but necessarily finite) systems.

\textit{Model and formalism.}---We consider non-equilibrium spin
dynamics generated by the Hamiltonian
\begin{equation}
\label{eq:hamiltonian}
H=\frac{J}{2}\sum_{i\neq j}\frac{\hat{\sigma}^z_i\hat{\sigma}^z_j}{|\bm{r}_i-\bm{r}_j|^{\alpha}},
\end{equation}
where indices $i,j$ label sites of a regular lattice, which are located
at positions $\bm{r}_{j}$.  Note that this Hamiltonian reduces, in the
limit $\alpha\rightarrow 0$, to the single-axis twisting model.  The simplest protocol for generating spin-squeezing from the Hamiltonian
in Eq.~\eqref{eq:hamiltonian} is to initiate all of the spins pointing
along a common direction in the $xy$ plane; without loss of generality
we choose the $x$-axis [Fig.~\ref{fig:squeezing_and_correlation_graphic}(a)].  The variance of the total spin measured along a direction $\cos(\psi)\hat{y}+\sin(\psi)\hat{z}$
perpendicular to the $x$-axis is given by
\begin{equation}
\Delta_{\psi}(t) ^2=\frac{1}{4}\sum_{i,j}\langle(\hat{\sigma}_i^y \cos\psi+\hat{\sigma}_i^z \sin\psi)(\hat{\sigma}_j^y \cos\psi+\hat{\sigma}_j^z \sin\psi)\rangle,\nonumber
\end{equation}
where we have taken advantage of the fact that all single-spin expectation values in the $yz$ plane vanish
by symmetry.  In the initial state the variance is isotropic and given by $\Delta_{\psi}(0)=\mathcal{N}/4$.  At any later time $t$, the variance depends on $\psi$. Minimization of the variance with respect to the angle $\psi$ at fixed $t$ can be carried out without explicitly evaluating the correlation functions involved \cite{ueda}, yielding the squeezing parameter  as a function of time \footnote{As
  an entanglement witness or a measure of metrological enhancement,
  $\xi$ should be normalized by twice the Bloch vector length rather than
  the number of particles.  However, the
  Bloch vector length at the optimal squeezing time approaches $\mathcal{N}/2$ in the large
  $\mathcal{N}$ limit, and thus its omission does not affect the
  derived scaling exponents.},
\begin{equation}
\label{eq:xi}
\xi(t)^2\equiv \mathcal{N}^{-1}\Delta^2_{\rm min}(t)=1+\mathcal{A}(t)-\sqrt{\mathcal{A}(t)^2+\mathcal{B}(t)^2},
\end{equation}
where
\begin{eqnarray}
\label{At1}
\mathcal{A}(t)&=&\frac{1}{2\mathcal{N}}\sum_{i,j}\left(\langle\hat{\sigma}^{y}_i\hat{\sigma}^{y}_j\rangle-\langle\hat{\sigma}_i^z\hat{\sigma}_j^z\rangle\right),\\
\label{Bt1}
\mathcal{B}(t)&=&\frac{1}{2\mathcal{N}}\sum_{i,j}\left
  (\langle\hat{\sigma}_i^{y}\hat{\sigma}_j^{z}\rangle+\langle\hat{\sigma}_i^z\hat{\sigma}_j^y\rangle\right).
\end{eqnarray}

\begin{figure}[!tbp]
\begin{center}
\includegraphics[width = 1.0 \columnwidth]{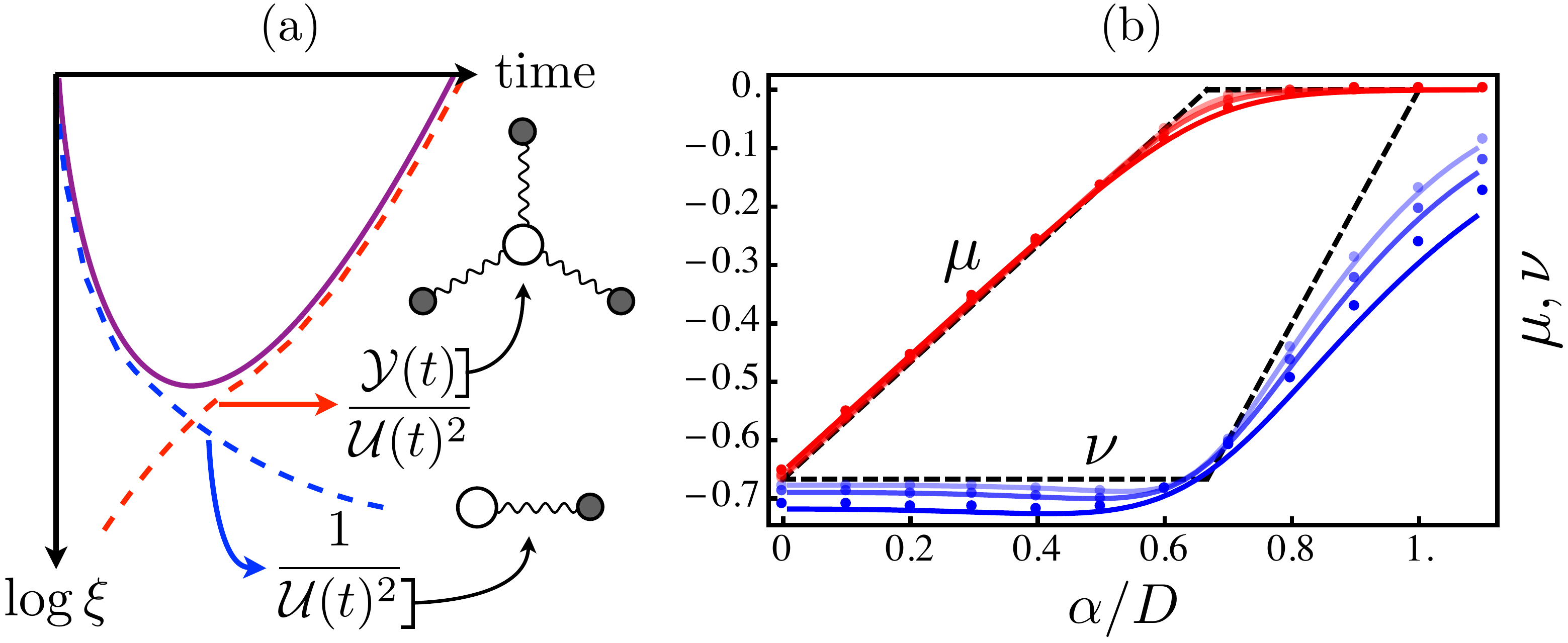}
\caption{(Color online) (a) Schematic plot of Eq.~\eqref{eq:answer},
  demonstrating how the optimal squeezing results from a competition
  of two different diagrams. (b) Scaling of spin-squeezing with the interaction exponent
  $\alpha$.  The blue (red) points show the scaling $\nu$ ($\mu$) for three different values of
  $\mathcal{N}=10^3,10^4,10^5$ (from darker to lighter shades), calculated using \eref{eq:Bt2}.  The
  solid blue (red) lines are the corresponding results obtained from \eref{eq:answer}. The
  black-dashed lines are analytic results, valid for $\mathcal{N}\rightarrow\infty$, and reported in the table on the fourth page of the manuscript.}
\label{fig:numerical_scaling}
\end{center}
\end{figure}

In general $\xi(t)$ will decrease with time (starting at $\xi(0)=1$), obtaining a minimum value before eventually growing large. We characterize the time $\tilde{t}$ at which $\xi$ is minimized,
and the squeezing $\tilde{\xi}=\xi(\,\tilde{t}\,)$ achieved at that time,
by how they scale with the system size $\mathcal{N}$ in the $\mathcal{N}\rightarrow\infty$ limit \footnote{These
 scaling exponents are defined formally by
  $\nu=\lim_{\mathcal{N}\rightarrow\infty}d
    \log\tilde{\xi}^2/d\log\mathcal{N}$, $\mu=\lim_{\mathcal{N}\rightarrow\infty}d\log\tilde{t}/d\log\mathcal{N}$.  These derivatives, evaluated
  at finite $\mathcal{N}$, determine the curves in
  Fig.~\ref{fig:numerical_scaling}.  The arbitrary choice
of $\xi^2$ instead of $\xi$ in the definition of $\nu$ is made so that
$\mu$ and $\nu$ are equal at $\alpha=0$.},
\begin{equation}
\tilde{\xi}^2\sim\mathcal{N}^{\nu}~~~~{\rm and}~~~~\tilde{t}\sim\mathcal{N}^{\mu}.
\end{equation}
For infinite-range interactions ($\alpha=0$), the behavior of the system
simplifies because the dynamics is constrained to a
small set of collective states, known as the Dicke manifold for its role in the theory of superradiance \cite{Dicke1954}.  Equation \eqref{eq:xi} then takes a simple closed form, and analysis of its asymptotic behavior yields $\nu=\mu=-2/3$ \cite{ueda}.  Note that $\nu=0$ and $\nu=-1$ correspond, respectively, to the standard quantum limit and the Heisenberg limit; the single-axis twisting model produces squeezed states that fall in between these two limits. 

\begin{figure}[h]
\begin{center}
\includegraphics[width = 0.99 \columnwidth]{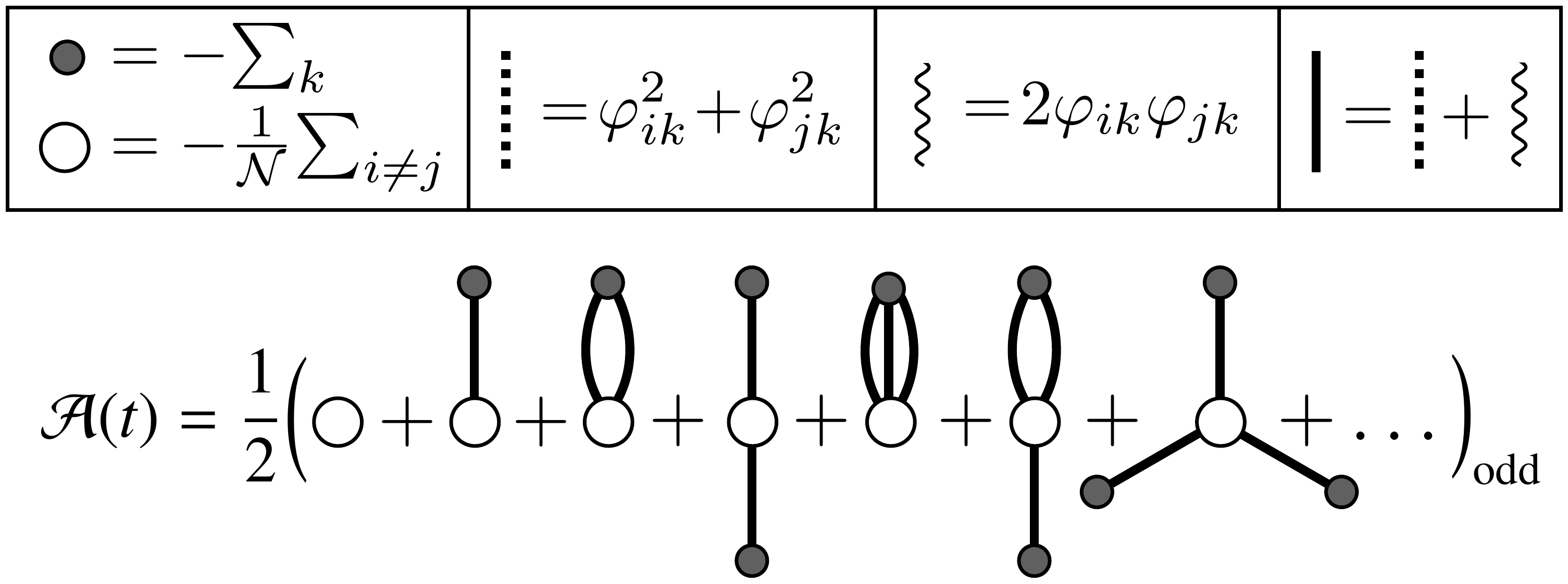}
\vspace{-0.2 cm}
\caption{Expansion of Eq.~\eqref{eq:Bt2} [see also Eq.~\eqref{eq:A_series}].
  Each diagram has an associated combinatorial factor, which is
  obtained by multiplying the entire diagram by $1/m!$ for each
  group of $m$ identical vertices, and then multiplying each
  black vertex by $1/(2n)!$, where $n$ is the
  number of legs connected to that vertex.  The subscript ``odd''
  implies to keep only diagrams with an odd number of wavy legs.}
\label{fig:A_diagrams}
\end{center}
\vspace{-0.4 cm}
\end{figure}

\emph{Scaling of squeezing for $\alpha\neq0$.}---For interactions that decay with distance the dynamics is no longer constrained to the Dicke manifold.  However,
correlation functions can still be computed using the exact solutions of
Refs.\,\cite{fossfeig2013,kastner2012}.  We find
\begin{align}
\mathcal{A}(t)&=\frac{1}{4\mathcal{N}}\!\sum_{i\neq j}\!\Big(\prod_{k\neq
  i,j}\!\cos\big(\varphi_{ik}\!-\!\varphi_{jk}\big)-\!\!\prod_{k\neq
  i,j}\!\cos\big(\varphi_{ik}\!+\!\varphi_{jk}\big)\Big),\nonumber\\
\label{eq:Bt2}
\mathcal{B}(t)&=\frac{1}{\mathcal{N}}\!\sum_{i\neq j}\sin\big(\varphi_{ij}\big)\prod_{k\neq i, j}\cos\varphi_{ik},
\end{align}
where $\varphi_{ik}=2Jt|\bm{r}_i-\bm{r}_k|^{-\alpha}$.  To
proceed we will expand these functions as series in time and analyze, at each
order in time, the leading contributions in powers of the system
size $\mathcal{N}$.  To this end, it is helpful to give the terms in these
expansions diagramatic representations.  For example,
Fig.~\ref{fig:A_diagrams} shows a schematic representation of the
series expansion for $\mathcal{A}(t)$.

\noindent These diagrams facilitate a sorting of contributions to $\mathcal{A}(t)$ and $\mathcal{B}(t)$ by powers of $\mathcal{N}$.  For example, in the case $\alpha=0$ it is
straightforward to see that tree diagrams---those for which each black vertex is connected by a single leg to the central white vertex---contribute dominantly (in powers of $\mathcal{N}$) at any order in time.

To obtain a valid expansion for the squeezing parameter $\xi(t)$ near $t=\tilde{t}$, we
assume that $\mathcal{A}(\tilde{t})\gg\mathcal{B}(\tilde{t})$, which can be verified at the end of the
calculation.  This assumption enables us to rewrite $\xi(t)$ as the following convergent series
\begin{equation}
\label{eq:xi_series}
\xi(t)^2=\!1-\frac{1}{2}\frac{\mathcal{B}(t)^2}{\mathcal{A}(t)}+\frac{1}{8}\frac{\mathcal{B}(t)^4}{\mathcal{A}(t)^3}-\dots
\end{equation}
Inspection of Eq.~\eqref{eq:xi_series} reveals that $\xi(t)\ll1$ requires $\delta(t)\equiv\mathcal{A}(t)-\frac{1}{2}\mathcal{B}(t)^2>0$
must satisfy $\delta(t)\ll\mathcal{A}(t),\frac{1}{2}\mathcal{B}(t)^2$.
Plugging $\mathcal{A}(t)=\frac{1}{2}\mathcal{B}(t)^2+\delta(t)$ into Eq.~\eqref{eq:xi_series} and working to lowest
nontrivial order in both $\delta(t)$ and $\mathcal{B}(t)^{-1}$, we then obtain
\begin{equation}
\label{eq:xi_approx}
\xi(t)^2\approx\frac{1+2\delta(t)}{\mathcal{B}(t)^{2}}.
\end{equation}
Our goal is now to understand the asymptotic behavior of Eq.~\eqref{eq:xi_approx} for large $\mathcal{N}$ by expanding the correlation functions involved diagramatically.

For $0<\alpha<D/2$, each black vertex attached by a single leg contributes at order
$t^2\mathcal{N}^{1-2\alpha/D}$ to a given diagram, as can be verified by converting the
sums involved into integrals \cite{SOM}.  Summations over black vertices with more than one
leg either converge or diverge less strongly with $\mathcal{N}$, and
therefore $\mathcal{A}(t)$ is dominated by tree diagrams. Factoring out all contributions from
wavy legs, we obtain
\begin{figure}[h!]
\begin{center}
\includegraphics[width = 1.0 \columnwidth]{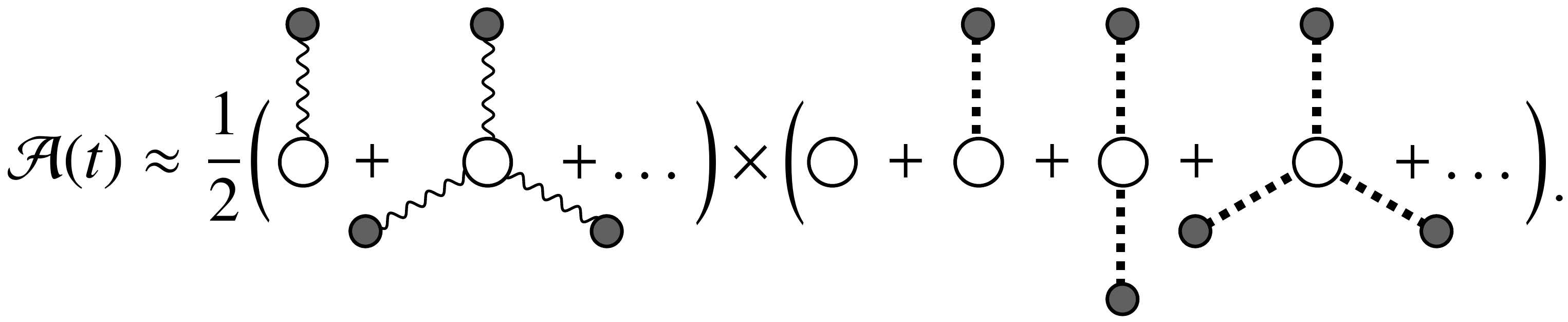}
\end{center}
\end{figure}
\vspace{-0.6 cm}

\noindent Here, ``$\approx$'' means ``equal to leading order in $\mathcal{N}$
at all orders in $t$'', and a product of two diagrams is carried out by joining
their white vertices. The two series in parentheses above can be resummed to give
\begin{equation}
\label{eq:Aapprox}
\mathcal{A}(t)\!\approx\!\frac{1}{2\mathcal{N}}\!\sum_{i\neq j}\sinh\!\big(\mathcal{V}(\bm{r}_i,\bm{r}_j) \big)\times \exp\big(-\sum_{k}(\varphi_{ik}^2+\varphi_{jk}^2)\big),
\end{equation}
where $\mathcal{V}(\bm{r}_i,\bm{r}_j)\equiv\sum_{k}\varphi_{ik}\varphi_{jk}$. In the supplemental material \cite{SOM}, we show that for a homogeneous system the terms with one wavy leg, which can be obtained from \eref{eq:Aapprox} by expanding the $\sinh$ to first order, coincide with the expansion of $\frac{1}{2}\mathcal{B}(t)^2$ to leading order in $\mathcal{N}$ at all orders in time (this also holds true for $D/2<\alpha<D$, even though non-tree diagrams must be included in this case).  Thus we have
\begin{figure}[!!h!!]
\begin{center}
\includegraphics[width = 1.0 \columnwidth]{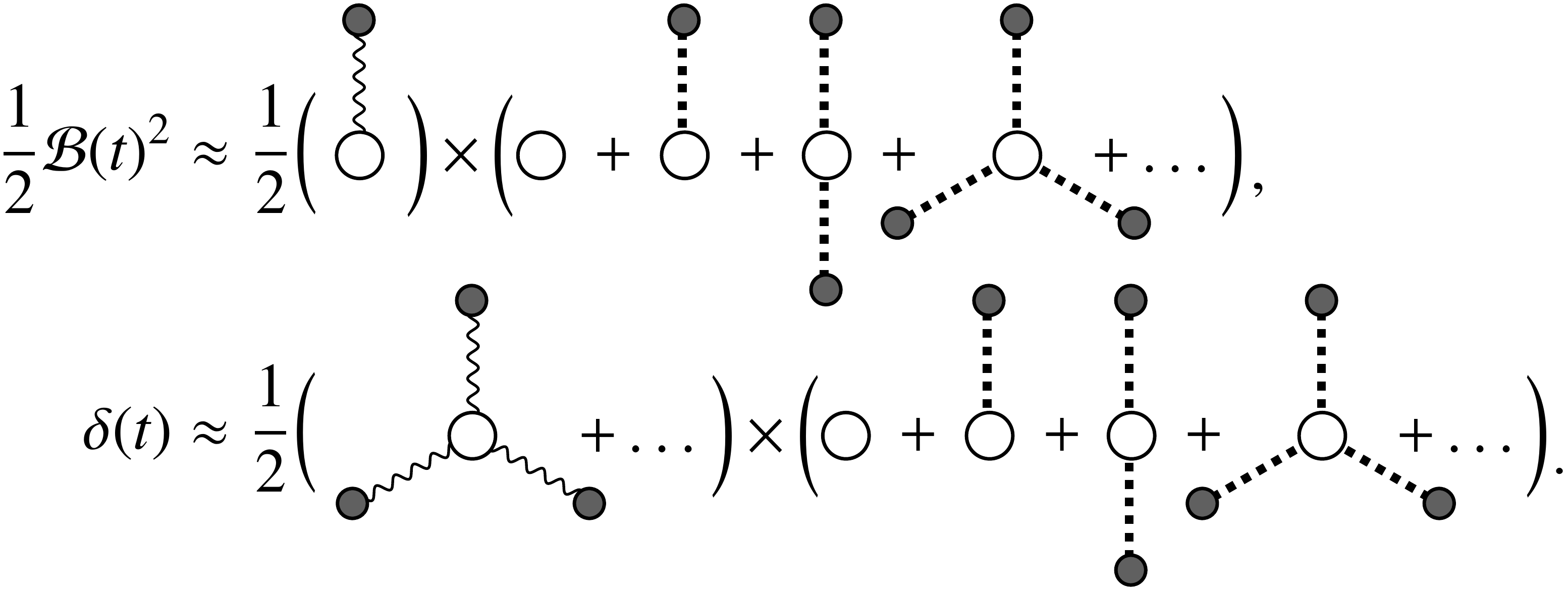}
\end{center}
\end{figure}
\vspace{-0.6 cm}

\noindent The equivalence between the set of diagrams with a single wavy leg
and $\frac{1}{2}\mathcal{B}(t)^2$ has a clear
physical origin:  In the expansion of $\mathcal{A}(t)$, these diagrams generate correlations between the $y$ components of two spins
at sites $i$ and $j$ through their mutual precession around the
$z$-projection of a single intermediate spin $k$.  On the other hand, that same diagram can be viewed
as the product of two independent processes that generate $yz$
correlations, either between the spins $i$ and $k$ or spins $j$ and
$k$, and therefore also provides the leading contribution in the
expansion of $\mathcal{B}(t)^2$.

In the exponent of \eref{eq:Aapprox}, we encounter sums that scale (at the time of optimal squeezing) as $\tilde{t}^2\times\mathcal{N}^{1-2\alpha/D}\sim\mathcal{N}^{2\mu+1-2\alpha/D}$.  These sums will be small in the large $\mathcal{N}$ limit as long as $\mu<\alpha/D-1/2$, in which case we set the exponential to unity; this condition will be assumed, and can be shown to be self-consistent at the end of the calculation.  Setting this exponential to unity amounts to dropping all diagrams containing dashed legs in $\mathcal{A}(t)$, $\mathcal{B}(t)^2$, and $\delta (t)$, and so at this level of approximation we have
\begin{align}
\mathcal{B}(t)^2\approx\frac{1}{\mathcal{N}}\sum_{i\neq j}\mathcal{V}(\bm{r}_i,\bm{r}_j)=\sum_{i\neq j}\varphi_{ik}\varphi_{jk}\approx\mathcal{U}(t)^2,
\end{align}
where $\mathcal{U}(t)\equiv\sum_{i}\varphi_{ik}$. Note that the equality results from homogeneity, and the second $\approx$ implies equality to leading order in $\mathcal{N}$. 
Still working under the assumption $\mu<\alpha/D-1/2$, one can check that at the time of optimal squeezing $\delta(t)$ is dominated by the lowest-order diagram (with three wavy legs), given by $\mathcal{Y}(t)=\frac{1}{\mathcal{N}}\frac{2^3}{3!}\sum_{i\neq j}\mathcal{V}(\bm{r}_i-\bm{r}_j)^3$, and therefore
\begin{align}
\label{eq:answer}
\xi^2(t)\approx\frac{1+\mathcal{Y}(t)}{\mathcal{U}(t)^2}.
\end{align}
The diagrams that represent $\mathcal{U}(t)$ and $\mathcal{Y}(t)$, and their contributions to the behavior of $\xi^2(t)$, are shown schematically in Fig.~\ref{fig:numerical_scaling}a.
Physically, $\mathcal{Y}(t)$ corresponds to a process where the $y$ components of two spins become
correlated via their mutual interaction with three other spins (as in Fig.~\ref{fig:squeezing_and_correlation_graphic}b). By converting sums into integrals and doing dimensional analysis, these diagrams can readily be seen to scale as $\mathcal{U}(t)\sim \mathcal{N}(t\mathcal{N}^{-\alpha/D})$ and $\mathcal{Y}(t)\sim \mathcal{N}^{4}(t\mathcal{N}^{-\alpha/D})^6$. Therefore, in terms of a rescaled time $\tau=t\mathcal{N}^{-\alpha/D}$, \eref{eq:answer} obtains the same form that it would in terms of the bare time $t$ at $\alpha=0$.  As an immediate consequence, the optimal squeezing exponent $\nu$ cannot depend on $\alpha$ and must take on its $\alpha=0$ value of $\nu=-2/3$. Similarly, we conclude that
$\tau\sim\mathcal{N}^{-2/3}$ (because this is how $t$ scales at $\alpha=0$), giving $\mu=-2/3+\alpha/D$.

Equation \eqref{eq:answer} actually remains valid as $\alpha$ exceed $D/2$, but the calculation leading to it becomes more difficult, in part because the expansion of $\mathcal{A}(t)$ is no longer dominated by tree diagrams.  A detailed analysis reveals that the leading order diagrams for $\mathcal{A}(t)$, which have one wavy leg,
continue to agree with $\frac{1}{2}\mathcal{B}(t)^2$ to leading order in $\mathcal{N}$ at all orders in time  \cite{SOM}.  In addition, $\mathcal{Y}(t)$ and $\mathcal{U}(t)^2$ continue to give the dominant contributions to $\delta(t)$ and $\mathcal{B}(t)^2$, respectively, as long as the diagram $\mathcal{Y}(t)$ diverges with $\mathcal{N}$, which persists until $\alpha=2D/3$.  Thus the result $(\nu,\mu)=(-2/3,\alpha/D-2/3)$ holds for $0\leq\alpha<2D/3$. For $2D/3<\alpha<D$, the numerator in Eq.~\eqref{eq:answer} no longer
diverges with the system size, while the denominator scales as
$t^2\mathcal{N}^{2-2\alpha/D}$.  The system-size dependence therefore
factors out, giving $\xi^2\sim\mathcal{N}^{2(\alpha/D-1)}$, causing the optimal squeezing to occur at a time that is
independent of $\mathcal{N}$.  For $\alpha>D$ all diagrams are
convergent in the large $\mathcal{N}$ limit, and neither $\tilde{t}$ nor $\tilde{\xi}$ scale with the system size, giving $\mu=\nu=0$.  These results can be summarized as follows,
\begin{figure}[!h]
\begin{center}
\includegraphics[width = 1.00 \columnwidth]{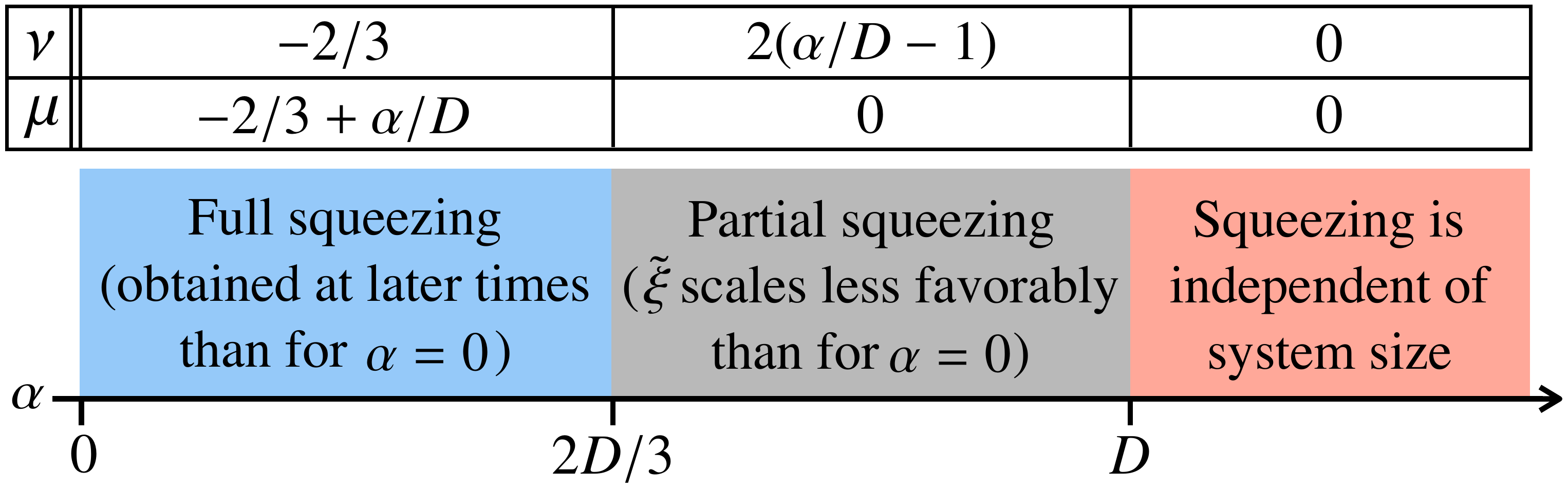}
\end{center}
\vspace{-0.6 cm}
\end{figure}
%
%
%

To verify Eq.~\eqref{eq:answer} and the associated scaling, we have computed $\mu$ and $\nu$ by direct numerical evaluation of Eq.~\eqref{eq:Bt2} for $\mathcal{N}=10^3,10^4,10^5$, and
plotted them as points in Fig.~\ref{fig:numerical_scaling}(b).  They can be seen
to converge toward the asymptotic scaling results (black dashed lines)
in the large system limit, and are increasingly well approximated by
Eq.~\eqref{eq:answer} for larger $\mathcal{N}$.



\textit{Effects of inhomogeneity.}---In the presence of spatially
decaying interactions, most experimental systems also lack spatial
homogeneity due to boundary effects (though exceptions certainly can
exist, e.g.~ring traps for ions \cite{TongcangLi2012}).  For sufficiently fast-decaying interactions, this lack of homogeneity is
only expected to be relevant near the edge of the system, and thus should not be important in the large $\mathcal{N}$ limit.
However, for interactions that are sufficiently long-ranged to
generate significant spin-squeezing ($\alpha<D$), edge effects cannot
in general be ignored (see for example Refs.~\cite{Gong13,Neyenhuis16}, in which boundary conditions in ion chains are shown to play an important role in local quench dynamics). In the absence of homogeneity, the cancellation of
leading order diagrams between $\mathcal{A}(t)$ and
$\frac{1}{2}\mathcal{B}(t)^2$ is imperfect, which results in a leading
  contribution to $\delta(t)$ determined by the variance of
  $\mathcal{U}_k(t)\equiv\sum_{i}\varphi_{ik}$ (which is now position dependent) 
\begin{equation}
\mathcal{X}(t)=\frac{1}{\mathcal{N}}\sum_{k}\mathcal{U}_k(t)^2-\Big(\frac{1}{\mathcal{N}}\sum_{k}\mathcal{U}_k(t)\Big)^2.
\end{equation}
For generic open boundary conditions and $\alpha<D$, this variance will
scale as $\mathcal{N}^{2-2\alpha/D}$ (the same as the lowest order
contribution to $\mathcal{B}(t)^2$), and vanishes for any finite system
as $\alpha\rightarrow 0$.  In the large system limit for \emph{any}
$\alpha>0$, we can thus approximate
\begin{equation}
\xi(t)^2\approx\frac{1+\mathcal{Y}(t)}{\mathcal{B}(t)^2}+\!\!\frac{\mathcal{X}(t)}{\mathcal{B}(t)^2}.
\end{equation}
Since the second term stays finite as
$\mathcal{N}\rightarrow\infty$, this expression leads to a minimum
squeezing parameter that is independent of system size.

\textit{Implications for experimental systems.}---These results shed
light on the importance of both translational invariance and
collective dynamics on spin-squeezing, and have implications for a
number of experimental systems where interactions are only
approximately independent of distance.  For example, in trapped ion
experiments, finite detuning from the center of mass mode results in
interactions that decay as a weak power law in space \cite{Britton12,islam13}.
\begin{figure}[!t]
\begin{center}
\includegraphics[width = 0.99 \columnwidth]{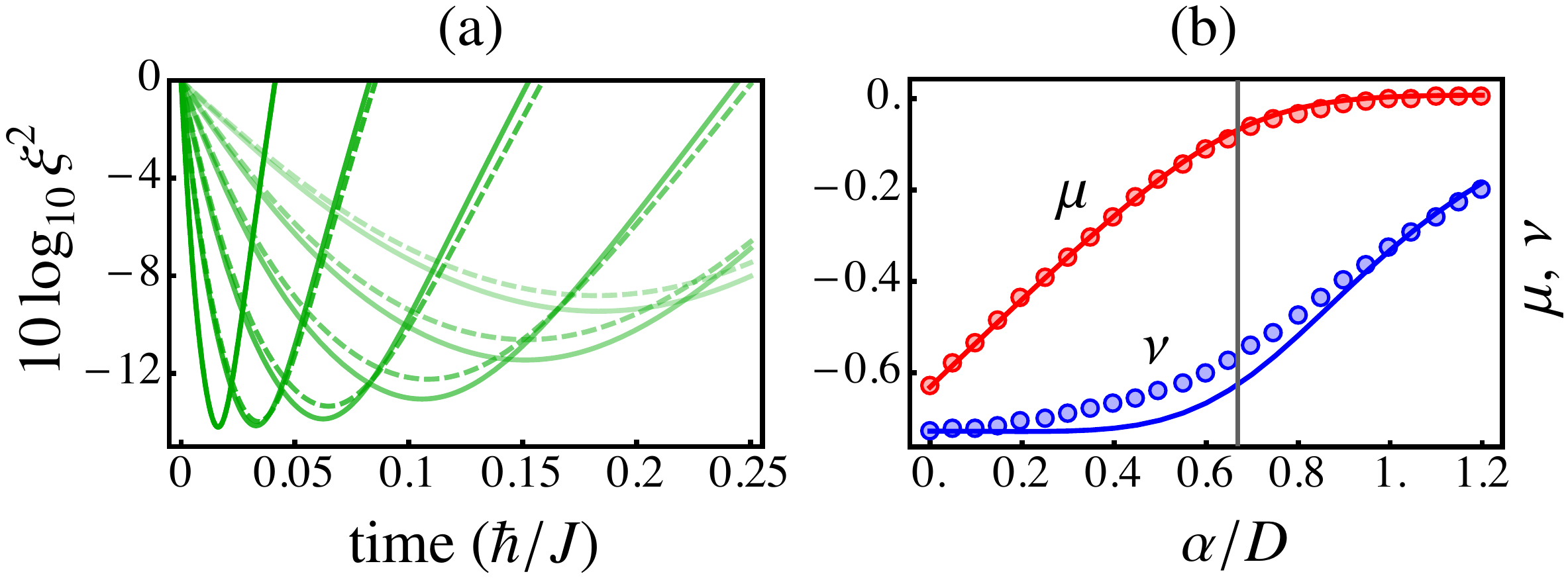}
\caption{(Color online) (a) Squeezing parameter as a
  function of time for a 1D chain with $200$ spins, for a variety of
  exponents $\alpha$ ranging from $0$ to $D=1$ (darker to lighter).
  Solid curves are for periodic boundary conditions, while dashed
  curves are for open boundary conditions.  (b) Scaling of squeezing with system size, again evaluated for a $200$-spin 1D chain.  The blue (red) line shows $\nu$ ($\mu$) for periodic boundary conditions, while
  the blue (red) disks show $\nu$ ($\mu$) for open
  boundary conditions.}
\label{fig:translational_invariance}
\end{center}
\vspace{-0.8 cm}
\end{figure}
For systems with open boundary conditions, our results demonstrate
that even arbitrarily slow-decaying interactions cause the maximum achievable
squeezing to be independent of system size in the large
$\mathcal{N}$ limit.  However, because the coefficient preceding
$\mathcal{N}^{2-2\alpha/D}$ in $\mathcal{X}(t)$ can be very small
($\mathcal{U}_k(t)$ does not fluctuate very much in space for
very long-range interactions), Eq.~\eqref{eq:answer} and the pursuant
scaling arguments are still relevant on a qualitative level for characterizing
spin squeezing with open boundary conditions, as long
as the system size is not too large.  In Fig.~\ref{fig:translational_invariance}(a) we show that the squeezing for
open and closed boundary conditions is not actually that different for
a linear chain with only hundreds of spins.  Moreover, in
Fig. \ref{fig:translational_invariance}(b) we see that the scaling behavior predicted for periodic boundary
conditions---e.g. squeezing that is weakly dependent on $\alpha$ for
$\alpha<2D/3$ and then falls off continuously, eventually vanishing for
$\alpha>D$---remains qualitatively correct for systems with open
boundary conditions.

\textit{Outlook.}---Entanglement generation via
spin squeezing has been shown to be robust against excursions outside of the Dicke (collective-spin) manifold.  For large systems the spin squeezing parameter is,
however, sensitive to the existence (or lack thereof) of
boundary-induced inhomogeneities.  It seems very likely that this sensitivity
indicates a failure of the squeezing parameter---which is explicitly
designed to capture the entanglement of permutationally (and therefore translationally) invariant spin
systems---to properly capture the entanglement generated in
inhomogeneous systems.  From a more fundamental standpoint, it would therefore
be worthwhile to explore whether the scaling with system size of quantum Fisher information \cite{PhysRevLett.72.3439,PhysRevA.87.032324}, entanglement depth
\cite{Sorensen2000}, or any other metrics of entanglement,
persists for spatially decaying interactions, even in
geometries where the scaling of the spin-squeezing parameter does not.

We are grateful to K.\,R.\,A.\,Hazzard, A.\,M.\,Rey, and  R.\,M.\,Wilson for
helpful discussions.  A.\,V.\,G.\ and Z.-X.\,G. acknowledge support from NSF QIS, ARO, ARO MURI, AFOSR, ARL CDQI, and NSF PFC at JQI.


%


\begin{widetext}

\renewcommand{\thesection}{S\arabic{section}} 
\renewcommand{\theequation}{S\arabic{equation}}
\renewcommand{\thefigure}{S\arabic{figure}}
\setcounter{equation}{0}
\setcounter{figure}{0}

\section{Supplemental material}

In this supplemental material, we describe in more detail how correlation functions can be expanded in terms of the diagrams described in the paper.  We also derive the scaling properties of different types of diagrams for various ranges of $\alpha$, and provide some additional details on the derivation leading from \eref{eq:xi_approx} to \eref{eq:answer} of the main text.

As a preliminary remark, we emphasize that the scaling of $\tilde{\xi}$ and $\tilde{t}$ in the large $\mathcal{N}$ limit cannot be understood by simply expanding $\xi(t)$ as
a power series in time.  The reason is that, though squeezing does occur at
short times (and thus $\tilde{t}$ is indeed a small parameter), the
existence of the large parameter $\mathcal{N}$ makes it impossible to
assign, a priori, preference to low-order terms in this expansion.  This is the fundamental reason why a careful expansion to high orders in time, which we carry out diagrammatically, is necessary.  As a simple (though still subtle) example,
consider the $\alpha=0$ case, which was first studied in Ref.\,\cite{ueda}.  There, one finds that while $\mathcal{N}t^2$ is a suitable
small parameter, $\mathcal{N}t$ actually becomes large (in the
large $\mathcal{N}$ limit) at the
time of optimal squeezing, as can be immediately verified by plugging in
the scaling $\tilde{t}\sim\mathcal{N}^{-2/3}$.  Therefore, a simultaneous
expansion in both $\varepsilon=\mathcal{N}t^2$ \emph{and}
$\delta=1/(\mathcal{N}t)$ is necessary to capture squeezing dynamics
at low order.  That such an expansion is not necessarily (and
indeed turns out not to be) analytic in time confirms our initial statement that low orders in short-time perturbation theory
will fail to describe spin squeezing.  It is nevertheless useful to expand all two-spin correlation functions as
series in time, and study the leading order contributions, in powers
of the system size $\mathcal{N}$, at each order.  Understanding this
scaling to all orders in time will allow us to identify suitable small
parameters in which to expand $\xi$, and to thereby identify its
scaling with $\mathcal{N}$ for non all-to-all interactions.

\subsection{Series expansion of correlation functions}

To begin, we can expand the correlation functions in Eq.~\eqref{eq:Bt2} of the manuscript.  Of particular importance is the expansion for $\mathcal{A}(t)$; defining $\Upsilon_{k}\equiv(\varphi_{ik}+\varphi_{jk})^2$, we find

\begin{align}
\label{eq:A_series}
\mathcal{A}(t)&=\frac{1}{4\mathcal{N}}\!\sum_{i\neq j}\!\Big(\prod_{k\neq
  i,j}\!\cos\big(\varphi_{ik}\!-\!\varphi_{jk}\big)-\!\!\prod_{k\neq
  i,j}\!\cos\big(\varphi_{ik}\!+\!\varphi_{jk}\big)\Big)\\
&=-\frac{1}{2\mathcal{N}}\sum_{i\neq
  j}\left(1-\sum_{k}\frac{\Upsilon_{k}}{2!}+\sum_{k}\frac{\Upsilon_{k}^2}{4!}+\frac{1}{2!}\sum_{k\neq
    l}\frac{\Upsilon_{k}}{2!}\frac{\Upsilon_{l}}{2!}-\sum_{k}\frac{\Upsilon_k^3}{6!}-\sum_{k\neq
    l}\frac{\Upsilon_k^{\phantom 2}}{2!}\frac{\Upsilon_l^2}{4!}-\frac{1}{3!}\sum_{k\neq
    l\neq
    m}\frac{\Upsilon_k}{2!}\frac{\Upsilon_l}{2!}\frac{\Upsilon_m}{2!}+\dots\right)_{\rm
odd}\nonumber
\end{align}
\vspace{-0.5 cm}
\begin{figure}[htbp]
\begin{center}
\includegraphics[width = 0.92 \columnwidth]{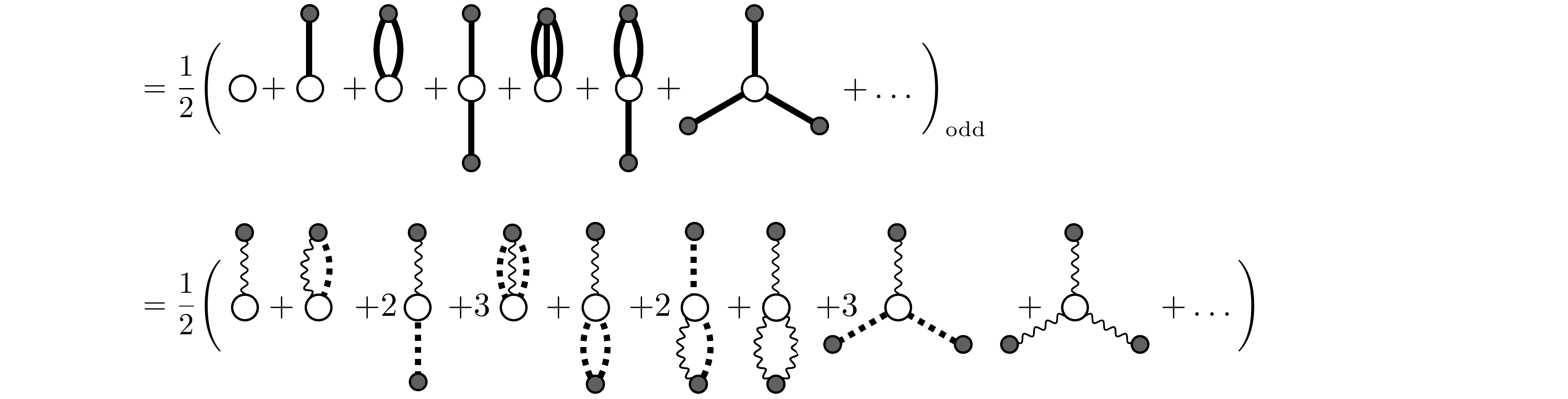}
\label{fig:supp_diagrams_1}
\end{center}
\end{figure}
\vspace{-0.2 cm}

\noindent In the second line, the subscript ``odd'' implies that, when expanding out a product of multiple $\Upsilon$'s, we should only keep terms that have an odd number of cross terms of the form $\varphi_{ik}\varphi_{jk}$.  In the third line, in light of the rules in \fref{fig:A_diagrams} of the manuscript, the subscript implies that we should only keep diagrams with an odd number of wavy legs.

\vspace{-0.4cm}

\subsection{Analysis of the leading order contributions}

In the case of $\alpha=0$, each black vertex contributes a power of
$\mathcal{N}$ to a diagram.  Given that each leg contributes a factor of $t^2$, the leading diagrams at each order in
time are all those with the same number of legs and black vertices.  In
the case of $m$ legs, such diagrams are proportional to $t^{2m}\mathcal{N}^{1+m}$.
For $0<\alpha<D/2$, each vertex with a single leg contributes a factor of
$t^2\mathcal{N}^{1-2\alpha/D}$, while any vertex with more than one
leg either converges or diverges less strongly with $\mathcal{N}$.  This result is easy to check for a
dashed-line vertex by analyzing the sums
\begin{equation}
\sum_{k\neq i}\frac{1}{|\bm{r}_i-\bm{r}_k|^{2m\alpha}}\sim\left\{ \begin{array}{ll}
         \mathcal{N}^{1-2m\alpha/D} & \mbox{if}~2m\alpha<1;\\
       1&\mbox{if}~2m\alpha>1.\end{array} \right.
\end{equation} 
Thus for $\alpha<D/2$ a black vertex with a single dashed leg always diverges, and grows
faster with system size than a black vertex with more than one leg.  The
contribution of a vertex with a single wavy leg depends on the positions of sites $i$ and $j$, which are eventually summed over.  To analyze the
contribution from such vertices we must therefore
understand the behavior of
$\mathcal{V}(\bm{r}_i,\bm{r}_j)=\sum_{k}|\bm{r}_i-\bm{r}_k|^{-\alpha}|\bm{r}_k-\bm{r}_j|^{-\alpha}$.
This sum can be estimated by converting it to an integral; for example in 1D, using a continuous coordinate
rescaled by the system size $z=j/\mathcal{N}$, we would have
\begin{equation}
\label{eqSOM:integral_approx}
\mathcal{V}(\bm{r}_i,\bm{r}_j)\approx\mathcal{N}^{1-2\alpha}\int_0^1 \frac{dz}{|x-z|^{\alpha}|z-y|^{\alpha}}.
\end{equation}
This integral converges for all $x\neq y$ whenever $\alpha<D$ (the
singularities at $z=x,y$ are integrable), and thus the overall scaling
of $\mathcal{V}$ is determined entirely by the
prefactor $\mathcal{N}^{1-2\alpha}$.  For $\alpha<D/2$ the
integral remains finite even as $x\rightarrow y$, which implies that $\mathcal{V}(\bm{r}_i,\bm{r}_j)$ remains
proportional to $\mathcal{N}^{1-2\alpha}$ in the large $\mathcal{N}$
limit even if the separations between sites $i$ and $j$ grows with the
system size.  Thus we find that $\mathcal{V}(\bm{r}_i,\bm{r}_j)\sim\mathcal{N}^{1-2\alpha}f(i,j)$,
where $f(i,j)$ is a function that stays finite in the limit of large
(i.e. on the order of the system size) values of
$|\bm{r}_i-\bm{r}_j|$.  The validity of the above derivation does not rely (except
in detail) on the lattice structure or dimension, and hence we find
more generally that
$\mathcal{V}(\bm{r}_i,\bm{r}_j)\sim\mathcal{N}^{1-2\alpha/D}f(i,j)$.
Because the function $f(i,j)$ tends to a finite value for extensively
large separations, the summation over $i$ and $j$ yields two powers of
the system size $\mathcal{N}$, and therefore the contribution of a
diagram with $m$ legs and vertices scales as $\mathcal{N}^{1+m(1-2\alpha/D)}$ regardless of the types of legs.

In light of the above discussion, and still restricting our attention
to the case $\alpha<D/2$, we can approximate the correlation function $\mathcal{A}(t)$ by keeping just
the leading order contributions in $\mathcal{N}$ at each order in
time:
\begin{figure}[!h]
\begin{center}
\includegraphics[width = 0.7 \columnwidth]{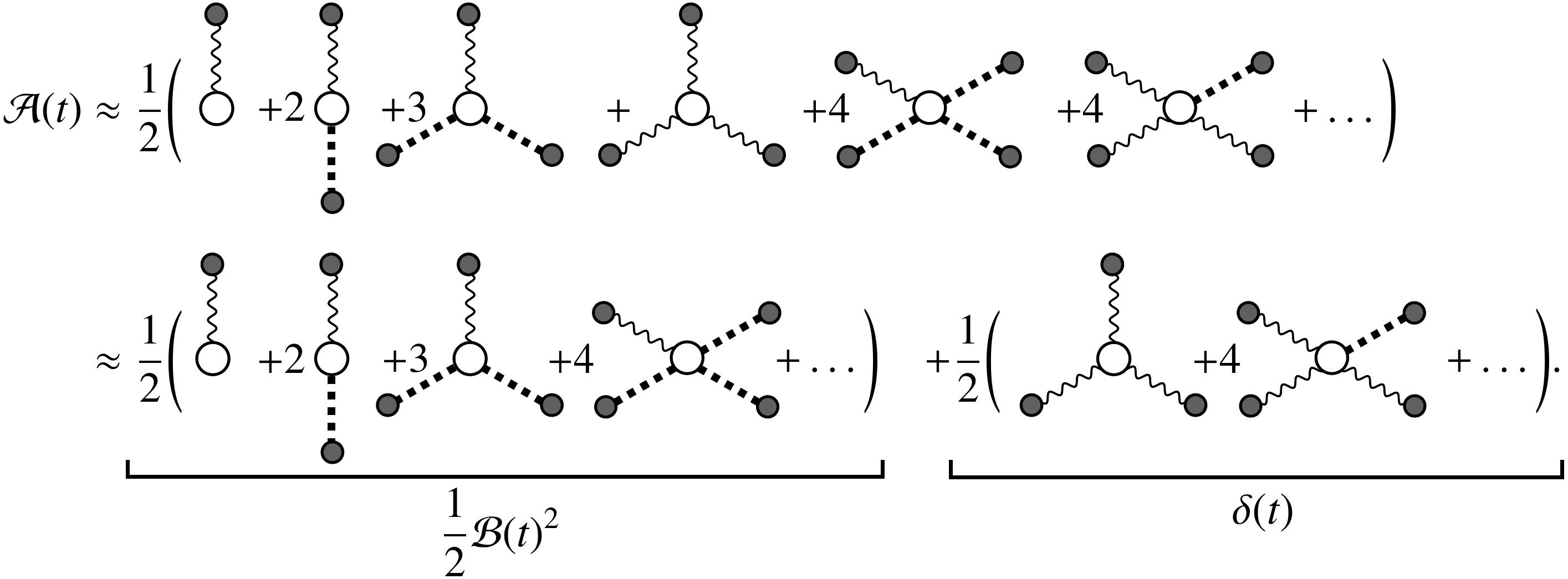}
\label{fig:supp_diagrams_2}
\end{center}
\end{figure}
\vspace{-0.6cm}

\noindent The set of diagrams with a single wavy leg can be simplified by factoring out the vertex
common to all diagrams, and resumming the remaining diagrams to give
\begin{align}
\label{eqSOM:a_vs_b_1}
\frac{1}{2}\frac{1}{\mathcal{N}}\sum_{i\neq
  j}\mathcal{V}(\bm{r}_i,\bm{r}_j)\exp[-\sum_{k\neq i,j}(\varphi_{ik}^2+\varphi_{jk}^2)]&=\frac{1}{2}\sum_{i\neq j}\Big(\varphi_{il}\exp(-\sum_{k\neq
  i,j}\varphi_{ik}^2)\Big)\Big(\varphi_{jl}\exp(-\sum_{k\neq
  i,j}\varphi_{jk}^2)\Big)\\
\label{eqSOM:a_vs_b_2}
&\approx\frac{1}{2}\mathcal{B}(t)^2,
\end{align}
where the $\approx$ means ``correct to leading order in
$\mathcal{N}$'' at each order in time.   The equality in
Eq.~\eqref{eqSOM:a_vs_b_1} can be verified by recalling the definition of $\mathcal{V}$ and exchanging summations:
\begin{equation}
\label{eqSOM:sum_exchange}
\sum_{i\neq j}\mathcal{V}(\bm{r}_i,\bm{r}_j)=\sum_{l}\sum_{i\neq j\neq
l}\frac{1}{|\bm{r}_i-\bm{r}_l|^{\alpha}|\bm{r}_l-\bm{r}_j|^{\alpha}}=\mathcal{N}\sum_{i\neq
j}\varphi_{il}\varphi_{jl}.
\end{equation}
Eq.~\eqref{eqSOM:a_vs_b_2} then follows immediately by expanding the
$\sin$ functions in the definition of $\mathcal{B}(t)$ to lowest order
(which is correct to leading order in $\mathcal{N}$).  This approximate equality is the
origin of the similarity between $\mathcal{A}(t)$ and
$\frac{1}{2}\mathcal{B}(t)^2$.  We can also resum the entire series
(including any number of wavy legs) to give
\begin{equation}
\mathcal{A}(t)\approx\frac{1}{2}\frac{1}{\mathcal{N}}\sum_{i\neq
  j}\sinh[\mathcal{V}(\bm{r}_i,\bm{r}_j)]\exp[-\sum_{k\neq i,j}(\varphi_{ik}^2+\varphi_{jk}^2)].
\end{equation}
Expanding the $\sinh$ function to first order gives the
$\frac{1}{2}\mathcal{B}(t)^2$ contribution, while the $3^{\rm rd}$
and higher order terms give $\delta(t)$.  Now that we have the desired series
representation for $\delta(t)$, and can see that the leading terms at
order $t^{2m}$ in time scale as $\mathcal{N}^{1+m(1-2\alpha/D)}$.  Truncating this series to lowest
($6^{\rm th}$) order in time and plugging it into Eq.~\eqref{eq:xi_approx} of
the manuscript leads immediately to Eq.~\eqref{eq:answer}.

The situation for $\alpha>D/2$ requires slightly more care.  First of
all, dashed legs ($\sim\sum_{k}|\bm{r}_i-\bm{r}_k|^{-2\alpha}$) now converge, and hence the diagrams with a single leg per vertex are not
parametrically larger (in $\mathcal{N}$) than diagrams with the same
number of legs but fewer
vertices---hence we cannot select a simple set of
dominant diagrams at a given order in time.  Similarly, for fixed $|\bm{r}_i-\bm{r}_j|$,
$\mathcal{V}(\bm{r}_i,\bm{r}_j)$ no longer diverges in the large
$\mathcal{N}$ limit, as suggested by the prefactor in Eq.~\eqref{eqSOM:integral_approx}.  Indeed, for a separation $|\bm{r}_i-\bm{r}_j|$
that grows proportionally to the system size,
$\mathcal{V}(\bm{r}_i,\bm{r}_j)$ actually tends to zero as a power law as $\mathcal{N}\rightarrow\infty$.  An immediate consequence is that the leading order
contributions to $\mathcal{A}(t)$ are diagrams with only a single wavy
leg, which suggests (incorrectly) that the leading contributions to
$\delta(t)$ might be of a different character than they were for $\alpha<D/2$.  However, these leading
order contributions are still precisely (i.e. to leading order in $\mathcal{N}$) canceled by the factor of
$\frac{1}{2}\mathcal{B}(t)^2$ in the definition of $\delta(t)$.  The cancelation of this more general class of diagrams, which includes those having one wavy
 leg and other dashed legs connected to the same vertex, is a consequence of the identity $\cos(a\pm
  b)=\cos(a)\cos(b)\mp\sin(a)\sin(b)$, employed inside the products of
equation Eq.~\eqref{eq:Bt2} \emph{before} the expansion.

To determine the next-to-leading-order contributions, we note that for
$\alpha>D/2$ the integral in Eq.~\eqref{eqSOM:integral_approx} has a power law divergence
$\sim|x-y|^{1-2\alpha}$ for small $|x-y|$.  In a diagram with $m$
wavy legs, the summation over $i$ and $j$ can be converted into an
integral
\begin{equation}
\sum_{i\neq j}\mathcal{V}(\bm{r}_i,\bm{r}_j)^{m}\approx\mathcal{N}^{1+m(1-2\alpha)}\int
dr\left(\int\frac{dz}{|z|^{\alpha}|z-r|^{\alpha}}\right)^m.
\end{equation}
Because the inner integral has a singularity $\sim r^{1-2\alpha}$, the
outer integral converges whenever $m(2\alpha-1)<1$, and so the overall
scaling of the digram is determined by the prefactor $\mathcal{N}^{1+m(1-2\alpha)}$.  For
$m(2\alpha-1)>1$ the outer integral gains a system-size dependence
(due to the non-integrable singularity at $r=0$)
$\sim\mathcal{N}^{m(2\alpha-1)-1}$, canceling the prefactor and causing
the diagram to converge in the large $\mathcal{N}$ limit.
From this discussion, it follows that progressively higher order diagrams converge first as $\alpha$ exceeds
$D/2$, and the last remaining divergent diagram contributing to
$\delta(t)$, namely $\mathcal{Y}(t)\sim\mathcal{N}^{4-6\alpha/D}$,
converges for $\alpha>2D/3$.

\vspace{0.5 cm}

\end{widetext}


\begin{thebibliography}{48}%
\makeatletter
\providecommand \@ifxundefined [1]{%
 \@ifx{#1\undefined}
}%
\providecommand \@ifnum [1]{%
 \ifnum #1\expandafter \@firstoftwo
 \else \expandafter \@secondoftwo
 \fi
}%
\providecommand \@ifx [1]{%
 \ifx #1\expandafter \@firstoftwo
 \else \expandafter \@secondoftwo
 \fi
}%
\providecommand \natexlab [1]{#1}%
\providecommand \enquote  [1]{``#1''}%
\providecommand \bibnamefont  [1]{#1}%
\providecommand \bibfnamefont [1]{#1}%
\providecommand \citenamefont [1]{#1}%
\providecommand \href@noop [0]{\@secondoftwo}%
\providecommand \href [0]{\begingroup \@sanitize@url \@href}%
\providecommand \@href[1]{\@@startlink{#1}\@@href}%
\providecommand \@@href[1]{\endgroup#1\@@endlink}%
\providecommand \@sanitize@url [0]{\catcode `\\12\catcode `\$12\catcode
  `\&12\catcode `\#12\catcode `\^12\catcode `\_12\catcode `\%12\relax}%
\providecommand \@@startlink[1]{}%
\providecommand \@@endlink[0]{}%
\providecommand \url  [0]{\begingroup\@sanitize@url \@url }%
\providecommand \@url [1]{\endgroup\@href {#1}{\urlprefix }}%
\providecommand \urlprefix  [0]{URL }%
\providecommand \Eprint [0]{\href }%
\providecommand \doibase [0]{http://dx.doi.org/}%
\providecommand \selectlanguage [0]{\@gobble}%
\providecommand \bibinfo  [0]{\@secondoftwo}%
\providecommand \bibfield  [0]{\@secondoftwo}%
\providecommand \translation [1]{[#1]}%
\providecommand \BibitemOpen [0]{}%
\providecommand \bibitemStop [0]{}%
\providecommand \bibitemNoStop [0]{.\EOS\space}%
\providecommand \EOS [0]{\spacefactor3000\relax}%
\providecommand \BibitemShut  [1]{\csname bibitem#1\endcsname}%
\let\auto@bib@innerbib\@empty
\bibitem [{\citenamefont {Lieb}\ and\ \citenamefont {Robinson}(1972)}]{lieb72}%
  \BibitemOpen
  \bibfield  {author} {\bibinfo {author} {\bibfnamefont {E.~H.}\ \bibnamefont
  {Lieb}}\ and\ \bibinfo {author} {\bibfnamefont {D.~W.}\ \bibnamefont
  {Robinson}},\ }\href@noop {} {\bibfield  {journal} {\bibinfo  {journal}
  {Comm. Math. Phys.}\ }\textbf {\bibinfo {volume} {28}},\ \bibinfo {pages}
  {251} (\bibinfo {year} {1972})}\BibitemShut {NoStop}%
\bibitem [{\citenamefont {Calabrese}\ and\ \citenamefont
  {Cardy}(2005)}]{Calabrese2005}%
  \BibitemOpen
  \bibfield  {author} {\bibinfo {author} {\bibfnamefont {P.}~\bibnamefont
  {Calabrese}}\ and\ \bibinfo {author} {\bibfnamefont {J.}~\bibnamefont
  {Cardy}},\ }\href {http://stacks.iop.org/1742-5468/2005/i=04/a=P04010}
  {\bibfield  {journal} {\bibinfo  {journal} {J. Stat. Mech.}\ }\textbf
  {\bibinfo {volume} {2005}},\ \bibinfo {pages} {P04010} (\bibinfo {year}
  {2005})}\BibitemShut {NoStop}%
\bibitem [{\citenamefont {Schachenmayer}\ \emph {et~al.}(2013)\citenamefont
  {Schachenmayer}, \citenamefont {Lanyon}, \citenamefont {Roos},\ and\
  \citenamefont {Daley}}]{schachenmayer13}%
  \BibitemOpen
  \bibfield  {author} {\bibinfo {author} {\bibfnamefont {J.}~\bibnamefont
  {Schachenmayer}}, \bibinfo {author} {\bibfnamefont {B.~P.}\ \bibnamefont
  {Lanyon}}, \bibinfo {author} {\bibfnamefont {C.~F.}\ \bibnamefont {Roos}}, \
  and\ \bibinfo {author} {\bibfnamefont {A.~J.}\ \bibnamefont {Daley}},\ }\href
  {\doibase 10.1103/PhysRevX.3.031015} {\bibfield  {journal} {\bibinfo
  {journal} {Phys. Rev. X}\ }\textbf {\bibinfo {volume} {3}},\ \bibinfo {pages}
  {031015} (\bibinfo {year} {2013})}\BibitemShut {NoStop}%
\bibitem [{\citenamefont {Dicke}(1954)}]{Dicke1954}%
  \BibitemOpen
  \bibfield  {author} {\bibinfo {author} {\bibfnamefont {R.~H.}\ \bibnamefont
  {Dicke}},\ }\href {\doibase 10.1103/PhysRev.93.99} {\bibfield  {journal}
  {\bibinfo  {journal} {Phys. Rev.}\ }\textbf {\bibinfo {volume} {93}},\
  \bibinfo {pages} {99} (\bibinfo {year} {1954})}\BibitemShut {NoStop}%
\bibitem [{\citenamefont {Hastings}\ and\ \citenamefont
  {Koma}(2006)}]{hastings05}%
  \BibitemOpen
  \bibfield  {author} {\bibinfo {author} {\bibfnamefont {M.~B.}\ \bibnamefont
  {Hastings}}\ and\ \bibinfo {author} {\bibfnamefont {T.}~\bibnamefont
  {Koma}},\ }\href {\doibase 10.1007/s00220-006-0030-4} {\bibfield  {journal}
  {\bibinfo  {journal} {Comm. Math. Phys.}\ }\textbf {\bibinfo {volume}
  {265}},\ \bibinfo {pages} {781} (\bibinfo {year} {2006})}\BibitemShut
  {NoStop}%
\bibitem [{\citenamefont {Hauke}\ and\ \citenamefont
  {Tagliacozzo}(2013)}]{hauke13}%
  \BibitemOpen
  \bibfield  {author} {\bibinfo {author} {\bibfnamefont {P.}~\bibnamefont
  {Hauke}}\ and\ \bibinfo {author} {\bibfnamefont {L.}~\bibnamefont
  {Tagliacozzo}},\ }\href {\doibase 10.1103/PhysRevLett.111.207202} {\bibfield
  {journal} {\bibinfo  {journal} {Phys. Rev. Lett.}\ }\textbf {\bibinfo
  {volume} {111}},\ \bibinfo {pages} {207202} (\bibinfo {year}
  {2013})}\BibitemShut {NoStop}%
\bibitem [{\citenamefont {Gong}\ \emph {et~al.}(2014)\citenamefont {Gong},
  \citenamefont {Foss-Feig}, \citenamefont {Michalakis},\ and\ \citenamefont
  {Gorshkov}}]{Gong14}%
  \BibitemOpen
  \bibfield  {author} {\bibinfo {author} {\bibfnamefont {Z.-X.}\ \bibnamefont
  {Gong}}, \bibinfo {author} {\bibfnamefont {M.}~\bibnamefont {Foss-Feig}},
  \bibinfo {author} {\bibfnamefont {S.}~\bibnamefont {Michalakis}}, \ and\
  \bibinfo {author} {\bibfnamefont {A.~V.}\ \bibnamefont {Gorshkov}},\ }\href
  {\doibase 10.1103/PhysRevLett.113.030602} {\bibfield  {journal} {\bibinfo
  {journal} {Phys. Rev. Lett.}\ }\textbf {\bibinfo {volume} {113}},\ \bibinfo
  {pages} {030602} (\bibinfo {year} {2014})}\BibitemShut {NoStop}%
\bibitem [{\citenamefont {Foss-Feig}\ \emph {et~al.}(2015)\citenamefont
  {Foss-Feig}, \citenamefont {Gong}, \citenamefont {Clark},\ and\ \citenamefont
  {Gorshkov}}]{fossfeig15}%
  \BibitemOpen
  \bibfield  {author} {\bibinfo {author} {\bibfnamefont {M.}~\bibnamefont
  {Foss-Feig}}, \bibinfo {author} {\bibfnamefont {Z.-X.}\ \bibnamefont {Gong}},
  \bibinfo {author} {\bibfnamefont {C.~W.}\ \bibnamefont {Clark}}, \ and\
  \bibinfo {author} {\bibfnamefont {A.~V.}\ \bibnamefont {Gorshkov}},\ }\href
  {\doibase 10.1103/PhysRevLett.114.157201} {\bibfield  {journal} {\bibinfo
  {journal} {Phys. Rev. Lett.}\ }\textbf {\bibinfo {volume} {114}},\ \bibinfo
  {pages} {157201} (\bibinfo {year} {2015})}\BibitemShut {NoStop}%
\bibitem [{\citenamefont {Eisert}\ \emph {et~al.}(2013)\citenamefont {Eisert},
  \citenamefont {van~den Worm}, \citenamefont {Manmana},\ and\ \citenamefont
  {Kastner}}]{eisert13}%
  \BibitemOpen
  \bibfield  {author} {\bibinfo {author} {\bibfnamefont {J.}~\bibnamefont
  {Eisert}}, \bibinfo {author} {\bibfnamefont {M.}~\bibnamefont {van~den
  Worm}}, \bibinfo {author} {\bibfnamefont {S.~R.}\ \bibnamefont {Manmana}}, \
  and\ \bibinfo {author} {\bibfnamefont {M.}~\bibnamefont {Kastner}},\ }\href
  {\doibase 10.1103/PhysRevLett.111.260401} {\bibfield  {journal} {\bibinfo
  {journal} {Phys. Rev. Lett.}\ }\textbf {\bibinfo {volume} {111}},\ \bibinfo
  {pages} {260401} (\bibinfo {year} {2013})}\BibitemShut {NoStop}%
\bibitem [{\citenamefont {Eldredge}\ \emph {et~al.}(2016)\citenamefont
  {Eldredge}, \citenamefont {Gong}, \citenamefont {Moosavian}, \citenamefont
  {Foss-Feig},\ and\ \citenamefont {Gorshkov}}]{Eldredge16}%
  \BibitemOpen
  \bibfield  {author} {\bibinfo {author} {\bibfnamefont {Z.}~\bibnamefont
  {Eldredge}}, \bibinfo {author} {\bibfnamefont {Z.-X.}\ \bibnamefont {Gong}},
  \bibinfo {author} {\bibfnamefont {A.~H.}\ \bibnamefont {Moosavian}}, \bibinfo
  {author} {\bibfnamefont {M.}~\bibnamefont {Foss-Feig}}, \ and\ \bibinfo
  {author} {\bibfnamefont {A.~V.}\ \bibnamefont {Gorshkov}},\ }\href
  {https://arxiv.org/abs/1612.02442} {\bibfield  {journal} {\bibinfo  {journal}
  {arXiv:1612.02442}\ } (\bibinfo {year} {2016})}\BibitemShut {NoStop}%
\bibitem [{\citenamefont {Yan}\ \emph {et~al.}(2013)\citenamefont {Yan},
  \citenamefont {Moses}, \citenamefont {Gadway}, \citenamefont {Covey},
  \citenamefont {Hazzard}, \citenamefont {Rey}, \citenamefont {Jin},\ and\
  \citenamefont {Ye}}]{yan13}%
  \BibitemOpen
  \bibfield  {author} {\bibinfo {author} {\bibfnamefont {B.}~\bibnamefont
  {Yan}}, \bibinfo {author} {\bibfnamefont {S.~A.}\ \bibnamefont {Moses}},
  \bibinfo {author} {\bibfnamefont {B.}~\bibnamefont {Gadway}}, \bibinfo
  {author} {\bibfnamefont {J.~P.}\ \bibnamefont {Covey}}, \bibinfo {author}
  {\bibfnamefont {K.~R.~A.}\ \bibnamefont {Hazzard}}, \bibinfo {author}
  {\bibfnamefont {A.~M.}\ \bibnamefont {Rey}}, \bibinfo {author} {\bibfnamefont
  {D.~S.}\ \bibnamefont {Jin}}, \ and\ \bibinfo {author} {\bibfnamefont
  {J.}~\bibnamefont {Ye}},\ }\href {http://dx.doi.org/10.1038/nature12483}
  {\bibfield  {journal} {\bibinfo  {journal} {Nature}\ }\textbf {\bibinfo
  {volume} {501}},\ \bibinfo {pages} {521} (\bibinfo {year}
  {2013})}\BibitemShut {NoStop}%
\bibitem [{\citenamefont {Richerme}\ \emph {et~al.}(2014)\citenamefont
  {Richerme}, \citenamefont {Gong}, \citenamefont {Lee}, \citenamefont {Senko},
  \citenamefont {Smith}, \citenamefont {Foss-Feig}, \citenamefont {Michalakis},
  \citenamefont {Gorshkov},\ and\ \citenamefont {Monroe}}]{richerme14}%
  \BibitemOpen
  \bibfield  {author} {\bibinfo {author} {\bibfnamefont {P.}~\bibnamefont
  {Richerme}}, \bibinfo {author} {\bibfnamefont {Z.-X.}\ \bibnamefont {Gong}},
  \bibinfo {author} {\bibfnamefont {A.}~\bibnamefont {Lee}}, \bibinfo {author}
  {\bibfnamefont {C.}~\bibnamefont {Senko}}, \bibinfo {author} {\bibfnamefont
  {J.}~\bibnamefont {Smith}}, \bibinfo {author} {\bibfnamefont
  {M.}~\bibnamefont {Foss-Feig}}, \bibinfo {author} {\bibfnamefont
  {S.}~\bibnamefont {Michalakis}}, \bibinfo {author} {\bibfnamefont {A.~V.}\
  \bibnamefont {Gorshkov}}, \ and\ \bibinfo {author} {\bibfnamefont
  {C.}~\bibnamefont {Monroe}},\ }\href {http://dx.doi.org/10.1038/nature13450}
  {\bibfield  {journal} {\bibinfo  {journal} {Nature}\ }\textbf {\bibinfo
  {volume} {511}},\ \bibinfo {pages} {198} (\bibinfo {year}
  {2014})}\BibitemShut {NoStop}%
\bibitem [{\citenamefont {Jurcevic}\ \emph {et~al.}(2014)\citenamefont
  {Jurcevic}, \citenamefont {Lanyon}, \citenamefont {Hauke}, \citenamefont
  {Hempel}, \citenamefont {Zoller}, \citenamefont {Blatt},\ and\ \citenamefont
  {Roos}}]{lanyon14}%
  \BibitemOpen
  \bibfield  {author} {\bibinfo {author} {\bibfnamefont {P.}~\bibnamefont
  {Jurcevic}}, \bibinfo {author} {\bibfnamefont {B.~P.}\ \bibnamefont
  {Lanyon}}, \bibinfo {author} {\bibfnamefont {P.}~\bibnamefont {Hauke}},
  \bibinfo {author} {\bibfnamefont {C.}~\bibnamefont {Hempel}}, \bibinfo
  {author} {\bibfnamefont {P.}~\bibnamefont {Zoller}}, \bibinfo {author}
  {\bibfnamefont {R.}~\bibnamefont {Blatt}}, \ and\ \bibinfo {author}
  {\bibfnamefont {C.~F.}\ \bibnamefont {Roos}},\ }\href
  {http://dx.doi.org/10.1038/nature13461} {\bibfield  {journal} {\bibinfo
  {journal} {Nature}\ }\textbf {\bibinfo {volume} {511}},\ \bibinfo {pages}
  {202} (\bibinfo {year} {2014})}\BibitemShut {NoStop}%
\bibitem [{\citenamefont {Kitagawa}\ and\ \citenamefont {Ueda}(1993)}]{ueda}%
  \BibitemOpen
  \bibfield  {author} {\bibinfo {author} {\bibfnamefont {M.}~\bibnamefont
  {Kitagawa}}\ and\ \bibinfo {author} {\bibfnamefont {M.}~\bibnamefont
  {Ueda}},\ }\href {\doibase 10.1103/PhysRevA.47.5138} {\bibfield  {journal}
  {\bibinfo  {journal} {Phys. Rev. A}\ }\textbf {\bibinfo {volume} {47}},\
  \bibinfo {pages} {5138} (\bibinfo {year} {1993})}\BibitemShut {NoStop}%
\bibitem [{\citenamefont {Wineland}\ \emph {et~al.}(1994)\citenamefont
  {Wineland}, \citenamefont {Bollinger}, \citenamefont {Itano},\ and\
  \citenamefont {Heinzen}}]{Wineland1994}%
  \BibitemOpen
  \bibfield  {author} {\bibinfo {author} {\bibfnamefont {D.~J.}\ \bibnamefont
  {Wineland}}, \bibinfo {author} {\bibfnamefont {J.~J.}\ \bibnamefont
  {Bollinger}}, \bibinfo {author} {\bibfnamefont {W.~M.}\ \bibnamefont
  {Itano}}, \ and\ \bibinfo {author} {\bibfnamefont {D.~J.}\ \bibnamefont
  {Heinzen}},\ }\href {\doibase 10.1103/PhysRevA.50.67} {\bibfield  {journal}
  {\bibinfo  {journal} {Phys. Rev. A}\ }\textbf {\bibinfo {volume} {50}},\
  \bibinfo {pages} {67} (\bibinfo {year} {1994})}\BibitemShut {NoStop}%
\bibitem [{\citenamefont {Esteve}\ \emph
  {et~al.}(2008{\natexlab{a}})\citenamefont {Esteve}, \citenamefont {Gross},
  \citenamefont {Weller}, \citenamefont {Giovanazzi},\ and\ \citenamefont
  {Oberthaler}}]{esteve}%
  \BibitemOpen
  \bibfield  {author} {\bibinfo {author} {\bibfnamefont {J.}~\bibnamefont
  {Esteve}}, \bibinfo {author} {\bibfnamefont {C.}~\bibnamefont {Gross}},
  \bibinfo {author} {\bibfnamefont {A.}~\bibnamefont {Weller}}, \bibinfo
  {author} {\bibfnamefont {S.}~\bibnamefont {Giovanazzi}}, \ and\ \bibinfo
  {author} {\bibfnamefont {M.~K.}\ \bibnamefont {Oberthaler}},\ }\href
  {http://dx.doi.org/10.1038/nature07332} {\bibfield  {journal} {\bibinfo
  {journal} {Nature}\ }\textbf {\bibinfo {volume} {455}},\ \bibinfo {pages}
  {1216} (\bibinfo {year} {2008}{\natexlab{a}})}\BibitemShut {NoStop}%
\bibitem [{\citenamefont {L\"{u}cke}\ \emph {et~al.}(2011)\citenamefont
  {L\"{u}cke}, \citenamefont {Scherer}, \citenamefont {Kruse}, \citenamefont
  {Pezz\'{e}}, \citenamefont {Deuretzbacher}, \citenamefont {Hyllus},
  \citenamefont {Topic}, \citenamefont {Peise}, \citenamefont {Ertmer},
  \citenamefont {Arlt}, \citenamefont {Santos}, \citenamefont {Smerzi},\ and\
  \citenamefont {Klempt}}]{Lucke2011}%
  \BibitemOpen
  \bibfield  {author} {\bibinfo {author} {\bibfnamefont {B.}~\bibnamefont
  {L\"{u}cke}}, \bibinfo {author} {\bibfnamefont {M.}~\bibnamefont {Scherer}},
  \bibinfo {author} {\bibfnamefont {J.}~\bibnamefont {Kruse}}, \bibinfo
  {author} {\bibfnamefont {L.}~\bibnamefont {Pezz\'{e}}}, \bibinfo {author}
  {\bibfnamefont {F.}~\bibnamefont {Deuretzbacher}}, \bibinfo {author}
  {\bibfnamefont {P.}~\bibnamefont {Hyllus}}, \bibinfo {author} {\bibfnamefont
  {O.}~\bibnamefont {Topic}}, \bibinfo {author} {\bibfnamefont
  {J.}~\bibnamefont {Peise}}, \bibinfo {author} {\bibfnamefont
  {W.}~\bibnamefont {Ertmer}}, \bibinfo {author} {\bibfnamefont
  {J.}~\bibnamefont {Arlt}}, \bibinfo {author} {\bibfnamefont {L.}~\bibnamefont
  {Santos}}, \bibinfo {author} {\bibfnamefont {A.}~\bibnamefont {Smerzi}}, \
  and\ \bibinfo {author} {\bibfnamefont {C.}~\bibnamefont {Klempt}},\ }\href
  {\doibase 10.1126/science.1208798} {\bibfield  {journal} {\bibinfo  {journal}
  {Science}\ }\textbf {\bibinfo {volume} {334}},\ \bibinfo {pages} {773}
  (\bibinfo {year} {2011})}\BibitemShut {NoStop}%
\bibitem [{\citenamefont {Chen}\ \emph {et~al.}(2011)\citenamefont {Chen},
  \citenamefont {Bohnet}, \citenamefont {Sankar}, \citenamefont {Dai},\ and\
  \citenamefont {Thompson}}]{Chen2011}%
  \BibitemOpen
  \bibfield  {author} {\bibinfo {author} {\bibfnamefont {Z.}~\bibnamefont
  {Chen}}, \bibinfo {author} {\bibfnamefont {J.~G.}\ \bibnamefont {Bohnet}},
  \bibinfo {author} {\bibfnamefont {S.~R.}\ \bibnamefont {Sankar}}, \bibinfo
  {author} {\bibfnamefont {J.}~\bibnamefont {Dai}}, \ and\ \bibinfo {author}
  {\bibfnamefont {J.~K.}\ \bibnamefont {Thompson}},\ }\href {\doibase
  10.1103/PhysRevLett.106.133601} {\bibfield  {journal} {\bibinfo  {journal}
  {Phys. Rev. Lett.}\ }\textbf {\bibinfo {volume} {106}},\ \bibinfo {pages}
  {133601} (\bibinfo {year} {2011})}\BibitemShut {NoStop}%
\bibitem [{\citenamefont {Foss-Feig}\ \emph {et~al.}(2013)\citenamefont
  {Foss-Feig}, \citenamefont {Hazzard}, \citenamefont {Bollinger},\ and\
  \citenamefont {Rey}}]{fossfeig2013}%
  \BibitemOpen
  \bibfield  {author} {\bibinfo {author} {\bibfnamefont {M.}~\bibnamefont
  {Foss-Feig}}, \bibinfo {author} {\bibfnamefont {K.~R.~A.}\ \bibnamefont
  {Hazzard}}, \bibinfo {author} {\bibfnamefont {J.~J.}\ \bibnamefont
  {Bollinger}}, \ and\ \bibinfo {author} {\bibfnamefont {A.~M.}\ \bibnamefont
  {Rey}},\ }\href {\doibase 10.1103/PhysRevA.87.042101} {\bibfield  {journal}
  {\bibinfo  {journal} {Phys. Rev. A}\ }\textbf {\bibinfo {volume} {87}},\
  \bibinfo {pages} {042101} (\bibinfo {year} {2013})}\BibitemShut {NoStop}%
\bibitem [{\citenamefont {Lee}\ \emph {et~al.}(2013)\citenamefont {Lee},
  \citenamefont {Gopalakrishnan},\ and\ \citenamefont {Lukin}}]{lee2013}%
  \BibitemOpen
  \bibfield  {author} {\bibinfo {author} {\bibfnamefont {T.~E.}\ \bibnamefont
  {Lee}}, \bibinfo {author} {\bibfnamefont {S.}~\bibnamefont {Gopalakrishnan}},
  \ and\ \bibinfo {author} {\bibfnamefont {M.~D.}\ \bibnamefont {Lukin}},\
  }\href {\doibase 10.1103/PhysRevLett.110.257204} {\bibfield  {journal}
  {\bibinfo  {journal} {Phys. Rev. Lett.}\ }\textbf {\bibinfo {volume} {110}},\
  \bibinfo {pages} {257204} (\bibinfo {year} {2013})}\BibitemShut {NoStop}%
\bibitem [{\citenamefont {Gil}\ \emph {et~al.}(2014)\citenamefont {Gil},
  \citenamefont {Mukherjee}, \citenamefont {Bridge}, \citenamefont {Jones},\
  and\ \citenamefont {Pohl}}]{Gil2013}%
  \BibitemOpen
  \bibfield  {author} {\bibinfo {author} {\bibfnamefont {L.~I.~R.}\
  \bibnamefont {Gil}}, \bibinfo {author} {\bibfnamefont {R.}~\bibnamefont
  {Mukherjee}}, \bibinfo {author} {\bibfnamefont {E.~M.}\ \bibnamefont
  {Bridge}}, \bibinfo {author} {\bibfnamefont {M.~P.~A.}\ \bibnamefont
  {Jones}}, \ and\ \bibinfo {author} {\bibfnamefont {T.}~\bibnamefont {Pohl}},\
  }\href {\doibase 10.1103/PhysRevLett.112.103601} {\bibfield  {journal}
  {\bibinfo  {journal} {Phys. Rev. Lett.}\ }\textbf {\bibinfo {volume} {112}},\
  \bibinfo {pages} {103601} (\bibinfo {year} {2014})}\BibitemShut {NoStop}%
\bibitem [{\citenamefont {Sorensen}\ \emph {et~al.}(2001)\citenamefont
  {Sorensen}, \citenamefont {Duan}, \citenamefont {Cirac},\ and\ \citenamefont
  {Zoller}}]{sorensen2001}%
  \BibitemOpen
  \bibfield  {author} {\bibinfo {author} {\bibfnamefont {A.}~\bibnamefont
  {Sorensen}}, \bibinfo {author} {\bibfnamefont {L.~M.}\ \bibnamefont {Duan}},
  \bibinfo {author} {\bibfnamefont {J.~I.}\ \bibnamefont {Cirac}}, \ and\
  \bibinfo {author} {\bibfnamefont {P.}~\bibnamefont {Zoller}},\ }\href
  {http://dx.doi.org/10.1038/35051038} {\bibfield  {journal} {\bibinfo
  {journal} {Nature}\ }\textbf {\bibinfo {volume} {409}},\ \bibinfo {pages}
  {63} (\bibinfo {year} {2001})}\BibitemShut {NoStop}%
\bibitem [{\citenamefont {S\o{}rensen}\ and\ \citenamefont
  {M\o{}lmer}(2001)}]{Sorensen2000}%
  \BibitemOpen
  \bibfield  {author} {\bibinfo {author} {\bibfnamefont {A.~S.}\ \bibnamefont
  {S\o{}rensen}}\ and\ \bibinfo {author} {\bibfnamefont {K.}~\bibnamefont
  {M\o{}lmer}},\ }\href {\doibase 10.1103/PhysRevLett.86.4431} {\bibfield
  {journal} {\bibinfo  {journal} {Phys. Rev. Lett.}\ }\textbf {\bibinfo
  {volume} {86}},\ \bibinfo {pages} {4431} (\bibinfo {year}
  {2001})}\BibitemShut {NoStop}%
\bibitem [{\citenamefont {Zhang}\ and\ \citenamefont {Duan}(2013)}]{Zhang2013}%
  \BibitemOpen
  \bibfield  {author} {\bibinfo {author} {\bibfnamefont {Z.}~\bibnamefont
  {Zhang}}\ and\ \bibinfo {author} {\bibfnamefont {L.-M.}\ \bibnamefont
  {Duan}},\ }\href {\doibase 10.1103/PhysRevLett.111.180401} {\bibfield
  {journal} {\bibinfo  {journal} {Phys. Rev. Lett.}\ }\textbf {\bibinfo
  {volume} {111}},\ \bibinfo {pages} {180401} (\bibinfo {year}
  {2013})}\BibitemShut {NoStop}%
\bibitem [{\citenamefont {Kuzmich}\ \emph {et~al.}(2000)\citenamefont
  {Kuzmich}, \citenamefont {Mandel},\ and\ \citenamefont
  {Bigelow}}]{PhysRevLett.85.1594}%
  \BibitemOpen
  \bibfield  {author} {\bibinfo {author} {\bibfnamefont {A.}~\bibnamefont
  {Kuzmich}}, \bibinfo {author} {\bibfnamefont {L.}~\bibnamefont {Mandel}}, \
  and\ \bibinfo {author} {\bibfnamefont {N.~P.}\ \bibnamefont {Bigelow}},\
  }\href {\doibase 10.1103/PhysRevLett.85.1594} {\bibfield  {journal} {\bibinfo
   {journal} {Phys. Rev. Lett.}\ }\textbf {\bibinfo {volume} {85}},\ \bibinfo
  {pages} {1594} (\bibinfo {year} {2000})}\BibitemShut {NoStop}%
\bibitem [{\citenamefont {Foss-Feig}\ \emph {et~al.}(2012)\citenamefont
  {Foss-Feig}, \citenamefont {Daley}, \citenamefont {Thompson},\ and\
  \citenamefont {Rey}}]{fossfeig2012}%
  \BibitemOpen
  \bibfield  {author} {\bibinfo {author} {\bibfnamefont {M.}~\bibnamefont
  {Foss-Feig}}, \bibinfo {author} {\bibfnamefont {A.~J.}\ \bibnamefont
  {Daley}}, \bibinfo {author} {\bibfnamefont {J.~K.}\ \bibnamefont {Thompson}},
  \ and\ \bibinfo {author} {\bibfnamefont {A.~M.}\ \bibnamefont {Rey}},\ }\href
  {\doibase 10.1103/PhysRevLett.109.230501} {\bibfield  {journal} {\bibinfo
  {journal} {Phys. Rev. Lett.}\ }\textbf {\bibinfo {volume} {109}},\ \bibinfo
  {pages} {230501} (\bibinfo {year} {2012})}\BibitemShut {NoStop}%
\bibitem [{\citenamefont {Dalla~Torre}\ \emph {et~al.}(2013)\citenamefont
  {Dalla~Torre}, \citenamefont {Otterbach}, \citenamefont {Demler},
  \citenamefont {Vuletic},\ and\ \citenamefont
  {Lukin}}]{PhysRevLett.110.120402}%
  \BibitemOpen
  \bibfield  {author} {\bibinfo {author} {\bibfnamefont {E.~G.}\ \bibnamefont
  {Dalla~Torre}}, \bibinfo {author} {\bibfnamefont {J.}~\bibnamefont
  {Otterbach}}, \bibinfo {author} {\bibfnamefont {E.}~\bibnamefont {Demler}},
  \bibinfo {author} {\bibfnamefont {V.}~\bibnamefont {Vuletic}}, \ and\
  \bibinfo {author} {\bibfnamefont {M.~D.}\ \bibnamefont {Lukin}},\ }\href
  {\doibase 10.1103/PhysRevLett.110.120402} {\bibfield  {journal} {\bibinfo
  {journal} {Phys. Rev. Lett.}\ }\textbf {\bibinfo {volume} {110}},\ \bibinfo
  {pages} {120402} (\bibinfo {year} {2013})}\BibitemShut {NoStop}%
\bibitem [{\citenamefont {Davis}\ \emph {et~al.}(2016)\citenamefont {Davis},
  \citenamefont {Bentsen},\ and\ \citenamefont
  {Schleier-Smith}}]{PhysRevLett.116.053601}%
  \BibitemOpen
  \bibfield  {author} {\bibinfo {author} {\bibfnamefont {E.}~\bibnamefont
  {Davis}}, \bibinfo {author} {\bibfnamefont {G.}~\bibnamefont {Bentsen}}, \
  and\ \bibinfo {author} {\bibfnamefont {M.}~\bibnamefont {Schleier-Smith}},\
  }\href {\doibase 10.1103/PhysRevLett.116.053601} {\bibfield  {journal}
  {\bibinfo  {journal} {Phys. Rev. Lett.}\ }\textbf {\bibinfo {volume} {116}},\
  \bibinfo {pages} {053601} (\bibinfo {year} {2016})}\BibitemShut {NoStop}%
\bibitem [{\citenamefont {Esteve}\ \emph
  {et~al.}(2008{\natexlab{b}})\citenamefont {Esteve}, \citenamefont {Gross},
  \citenamefont {Weller}, \citenamefont {Giovanazzi},\ and\ \citenamefont
  {Oberthaler}}]{gross}%
  \BibitemOpen
  \bibfield  {author} {\bibinfo {author} {\bibfnamefont {J.}~\bibnamefont
  {Esteve}}, \bibinfo {author} {\bibfnamefont {C.}~\bibnamefont {Gross}},
  \bibinfo {author} {\bibfnamefont {A.}~\bibnamefont {Weller}}, \bibinfo
  {author} {\bibfnamefont {S.}~\bibnamefont {Giovanazzi}}, \ and\ \bibinfo
  {author} {\bibfnamefont {M.~K.}\ \bibnamefont {Oberthaler}},\ }\href
  {http://dx.doi.org/10.1038/nature07332} {\bibfield  {journal} {\bibinfo
  {journal} {Nature}\ }\textbf {\bibinfo {volume} {455}},\ \bibinfo {pages}
  {1216} (\bibinfo {year} {2008}{\natexlab{b}})}\BibitemShut {NoStop}%
\bibitem [{\citenamefont {Monz}\ \emph {et~al.}(2011)\citenamefont {Monz},
  \citenamefont {Schindler}, \citenamefont {Barreiro}, \citenamefont {Chwalla},
  \citenamefont {Nigg}, \citenamefont {Coish}, \citenamefont {Harlander},
  \citenamefont {H\"ansel}, \citenamefont {Hennrich},\ and\ \citenamefont
  {Blatt}}]{Monz2011}%
  \BibitemOpen
  \bibfield  {author} {\bibinfo {author} {\bibfnamefont {T.}~\bibnamefont
  {Monz}}, \bibinfo {author} {\bibfnamefont {P.}~\bibnamefont {Schindler}},
  \bibinfo {author} {\bibfnamefont {J.~T.}\ \bibnamefont {Barreiro}}, \bibinfo
  {author} {\bibfnamefont {M.}~\bibnamefont {Chwalla}}, \bibinfo {author}
  {\bibfnamefont {D.}~\bibnamefont {Nigg}}, \bibinfo {author} {\bibfnamefont
  {W.~A.}\ \bibnamefont {Coish}}, \bibinfo {author} {\bibfnamefont
  {M.}~\bibnamefont {Harlander}}, \bibinfo {author} {\bibfnamefont
  {W.}~\bibnamefont {H\"ansel}}, \bibinfo {author} {\bibfnamefont
  {M.}~\bibnamefont {Hennrich}}, \ and\ \bibinfo {author} {\bibfnamefont
  {R.}~\bibnamefont {Blatt}},\ }\href {\doibase 10.1103/PhysRevLett.106.130506}
  {\bibfield  {journal} {\bibinfo  {journal} {Phys. Rev. Lett.}\ }\textbf
  {\bibinfo {volume} {106}},\ \bibinfo {pages} {130506} (\bibinfo {year}
  {2011})}\BibitemShut {NoStop}%
\bibitem [{\citenamefont {Bohnet}\ \emph {et~al.}(2014)\citenamefont {Bohnet},
  \citenamefont {Cox}, \citenamefont {Norcia}, \citenamefont {Weiner},
  \citenamefont {Chen},\ and\ \citenamefont {Thompson}}]{Bohnet2013}%
  \BibitemOpen
  \bibfield  {author} {\bibinfo {author} {\bibfnamefont {J.~G.}\ \bibnamefont
  {Bohnet}}, \bibinfo {author} {\bibfnamefont {K.~C.}\ \bibnamefont {Cox}},
  \bibinfo {author} {\bibfnamefont {M.~A.}\ \bibnamefont {Norcia}}, \bibinfo
  {author} {\bibfnamefont {J.~M.}\ \bibnamefont {Weiner}}, \bibinfo {author}
  {\bibfnamefont {Z.}~\bibnamefont {Chen}}, \ and\ \bibinfo {author}
  {\bibfnamefont {J.~K.}\ \bibnamefont {Thompson}},\ }\href
  {http://dx.doi.org/10.1038/nphoton.2014.151} {\bibfield  {journal} {\bibinfo
  {journal} {Nat. Photonics}\ }\textbf {\bibinfo {volume} {8}},\ \bibinfo
  {pages} {731} (\bibinfo {year} {2014})}\BibitemShut {NoStop}%
\bibitem [{\citenamefont {Hazzard}\ \emph {et~al.}(2014)\citenamefont
  {Hazzard}, \citenamefont {Gadway}, \citenamefont {Foss-Feig}, \citenamefont
  {Yan}, \citenamefont {Moses}, \citenamefont {Covey}, \citenamefont {Yao},
  \citenamefont {Lukin}, \citenamefont {Ye}, \citenamefont {Jin},\ and\
  \citenamefont {Rey}}]{Hazzard2014}%
  \BibitemOpen
  \bibfield  {author} {\bibinfo {author} {\bibfnamefont {K.~R.~A.}\
  \bibnamefont {Hazzard}}, \bibinfo {author} {\bibfnamefont {B.}~\bibnamefont
  {Gadway}}, \bibinfo {author} {\bibfnamefont {M.}~\bibnamefont {Foss-Feig}},
  \bibinfo {author} {\bibfnamefont {B.}~\bibnamefont {Yan}}, \bibinfo {author}
  {\bibfnamefont {S.~A.}\ \bibnamefont {Moses}}, \bibinfo {author}
  {\bibfnamefont {J.~P.}\ \bibnamefont {Covey}}, \bibinfo {author}
  {\bibfnamefont {N.~Y.}\ \bibnamefont {Yao}}, \bibinfo {author} {\bibfnamefont
  {M.~D.}\ \bibnamefont {Lukin}}, \bibinfo {author} {\bibfnamefont
  {J.}~\bibnamefont {Ye}}, \bibinfo {author} {\bibfnamefont {D.~S.}\
  \bibnamefont {Jin}}, \ and\ \bibinfo {author} {\bibfnamefont {A.~M.}\
  \bibnamefont {Rey}},\ }\href {\doibase 10.1103/PhysRevLett.113.195302}
  {\bibfield  {journal} {\bibinfo  {journal} {Phys. Rev. Lett.}\ }\textbf
  {\bibinfo {volume} {113}},\ \bibinfo {pages} {195302} (\bibinfo {year}
  {2014})}\BibitemShut {NoStop}%
\bibitem [{\citenamefont {de~Paz}\ \emph {et~al.}(2013)\citenamefont {de~Paz},
  \citenamefont {Sharma}, \citenamefont {Chotia}, \citenamefont {Mar\'echal},
  \citenamefont {Huckans}, \citenamefont {Pedri}, \citenamefont {Santos},
  \citenamefont {Gorceix}, \citenamefont {Vernac},\ and\ \citenamefont
  {Laburthe-Tolra}}]{dePaz2013}%
  \BibitemOpen
  \bibfield  {author} {\bibinfo {author} {\bibfnamefont {A.}~\bibnamefont
  {de~Paz}}, \bibinfo {author} {\bibfnamefont {A.}~\bibnamefont {Sharma}},
  \bibinfo {author} {\bibfnamefont {A.}~\bibnamefont {Chotia}}, \bibinfo
  {author} {\bibfnamefont {E.}~\bibnamefont {Mar\'echal}}, \bibinfo {author}
  {\bibfnamefont {J.~H.}\ \bibnamefont {Huckans}}, \bibinfo {author}
  {\bibfnamefont {P.}~\bibnamefont {Pedri}}, \bibinfo {author} {\bibfnamefont
  {L.}~\bibnamefont {Santos}}, \bibinfo {author} {\bibfnamefont
  {O.}~\bibnamefont {Gorceix}}, \bibinfo {author} {\bibfnamefont
  {L.}~\bibnamefont {Vernac}}, \ and\ \bibinfo {author} {\bibfnamefont
  {B.}~\bibnamefont {Laburthe-Tolra}},\ }\href {\doibase
  10.1103/PhysRevLett.111.185305} {\bibfield  {journal} {\bibinfo  {journal}
  {Phys. Rev. Lett.}\ }\textbf {\bibinfo {volume} {111}},\ \bibinfo {pages}
  {185305} (\bibinfo {year} {2013})}\BibitemShut {NoStop}%
\bibitem [{\citenamefont {Saffman}\ \emph {et~al.}(2010)\citenamefont
  {Saffman}, \citenamefont {Walker},\ and\ \citenamefont
  {M\o{}lmer}}]{RevModPhys.82.2313}%
  \BibitemOpen
  \bibfield  {author} {\bibinfo {author} {\bibfnamefont {M.}~\bibnamefont
  {Saffman}}, \bibinfo {author} {\bibfnamefont {T.~G.}\ \bibnamefont {Walker}},
  \ and\ \bibinfo {author} {\bibfnamefont {K.}~\bibnamefont {M\o{}lmer}},\
  }\href {\doibase 10.1103/RevModPhys.82.2313} {\bibfield  {journal} {\bibinfo
  {journal} {Rev. Mod. Phys.}\ }\textbf {\bibinfo {volume} {82}},\ \bibinfo
  {pages} {2313} (\bibinfo {year} {2010})}\BibitemShut {NoStop}%
\bibitem [{\citenamefont {Schau{\ss}}\ \emph {et~al.}(2012)\citenamefont
  {Schau{\ss}}, \citenamefont {Cheneau}, \citenamefont {Endres}, \citenamefont
  {Fukuhara}, \citenamefont {Hild}, \citenamefont {Omran}, \citenamefont
  {Pohl}, \citenamefont {Gross}, \citenamefont {Kuhr},\ and\ \citenamefont
  {Bloch}}]{nature11596}%
  \BibitemOpen
  \bibfield  {author} {\bibinfo {author} {\bibfnamefont {P.}~\bibnamefont
  {Schau{\ss}}}, \bibinfo {author} {\bibfnamefont {M.}~\bibnamefont {Cheneau}},
  \bibinfo {author} {\bibfnamefont {M.}~\bibnamefont {Endres}}, \bibinfo
  {author} {\bibfnamefont {T.}~\bibnamefont {Fukuhara}}, \bibinfo {author}
  {\bibfnamefont {S.}~\bibnamefont {Hild}}, \bibinfo {author} {\bibfnamefont
  {A.}~\bibnamefont {Omran}}, \bibinfo {author} {\bibfnamefont
  {T.}~\bibnamefont {Pohl}}, \bibinfo {author} {\bibfnamefont {C.}~\bibnamefont
  {Gross}}, \bibinfo {author} {\bibfnamefont {S.}~\bibnamefont {Kuhr}}, \ and\
  \bibinfo {author} {\bibfnamefont {I.}~\bibnamefont {Bloch}},\ }\href
  {http://dx.doi.org/10.1038/nature11596} {\bibfield  {journal} {\bibinfo
  {journal} {Nature}\ }\textbf {\bibinfo {volume} {491}},\ \bibinfo {pages}
  {87} (\bibinfo {year} {2012})}\BibitemShut {NoStop}%
\bibitem [{\citenamefont {Britton}\ \emph {et~al.}(2012)\citenamefont
  {Britton}, \citenamefont {Sawyer}, \citenamefont {Keith}, \citenamefont
  {Wang}, \citenamefont {Freericks}, \citenamefont {Uys}, \citenamefont
  {Biercuk},\ and\ \citenamefont {Bollinger}}]{Britton12}%
  \BibitemOpen
  \bibfield  {author} {\bibinfo {author} {\bibfnamefont {J.~W.}\ \bibnamefont
  {Britton}}, \bibinfo {author} {\bibfnamefont {B.~C.}\ \bibnamefont {Sawyer}},
  \bibinfo {author} {\bibfnamefont {A.~C.}\ \bibnamefont {Keith}}, \bibinfo
  {author} {\bibfnamefont {C.~C.~J.}\ \bibnamefont {Wang}}, \bibinfo {author}
  {\bibfnamefont {J.~K.}\ \bibnamefont {Freericks}}, \bibinfo {author}
  {\bibfnamefont {H.}~\bibnamefont {Uys}}, \bibinfo {author} {\bibfnamefont
  {M.~J.}\ \bibnamefont {Biercuk}}, \ and\ \bibinfo {author} {\bibfnamefont
  {J.~J.}\ \bibnamefont {Bollinger}},\ }\href
  {http://dx.doi.org/10.1038/nature10981} {\bibfield  {journal} {\bibinfo
  {journal} {Nature}\ }\textbf {\bibinfo {volume} {484}},\ \bibinfo {pages}
  {489} (\bibinfo {year} {2012})}\BibitemShut {NoStop}%
\bibitem [{\citenamefont {Bohnet}\ \emph {et~al.}(2016)\citenamefont {Bohnet},
  \citenamefont {Sawyer}, \citenamefont {Britton}, \citenamefont {Wall},
  \citenamefont {Rey}, \citenamefont {Foss-Feig},\ and\ \citenamefont
  {Bollinger}}]{Bohnet16}%
  \BibitemOpen
  \bibfield  {author} {\bibinfo {author} {\bibfnamefont {J.~G.}\ \bibnamefont
  {Bohnet}}, \bibinfo {author} {\bibfnamefont {B.~C.}\ \bibnamefont {Sawyer}},
  \bibinfo {author} {\bibfnamefont {J.~W.}\ \bibnamefont {Britton}}, \bibinfo
  {author} {\bibfnamefont {M.~L.}\ \bibnamefont {Wall}}, \bibinfo {author}
  {\bibfnamefont {A.~M.}\ \bibnamefont {Rey}}, \bibinfo {author} {\bibfnamefont
  {M.}~\bibnamefont {Foss-Feig}}, \ and\ \bibinfo {author} {\bibfnamefont
  {J.~J.}\ \bibnamefont {Bollinger}},\ }\href {\doibase
  10.1126/science.aad9958} {\bibfield  {journal} {\bibinfo  {journal}
  {Science}\ }\textbf {\bibinfo {volume} {352}},\ \bibinfo {pages} {1297}
  (\bibinfo {year} {2016})}\BibitemShut {NoStop}%
\bibitem [{\citenamefont {Martin}\ \emph {et~al.}(2013)\citenamefont {Martin},
  \citenamefont {Bishof}, \citenamefont {Swallows}, \citenamefont {Zhang},
  \citenamefont {Benko}, \citenamefont {von Stecher}, \citenamefont {Gorshkov},
  \citenamefont {Rey},\ and\ \citenamefont {Ye}}]{Martin2013}%
  \BibitemOpen
  \bibfield  {author} {\bibinfo {author} {\bibfnamefont {M.~J.}\ \bibnamefont
  {Martin}}, \bibinfo {author} {\bibfnamefont {M.}~\bibnamefont {Bishof}},
  \bibinfo {author} {\bibfnamefont {M.~D.}\ \bibnamefont {Swallows}}, \bibinfo
  {author} {\bibfnamefont {X.}~\bibnamefont {Zhang}}, \bibinfo {author}
  {\bibfnamefont {C.}~\bibnamefont {Benko}}, \bibinfo {author} {\bibfnamefont
  {J.}~\bibnamefont {von Stecher}}, \bibinfo {author} {\bibfnamefont {A.~V.}\
  \bibnamefont {Gorshkov}}, \bibinfo {author} {\bibfnamefont {A.~M.}\
  \bibnamefont {Rey}}, \ and\ \bibinfo {author} {\bibfnamefont
  {J.}~\bibnamefont {Ye}},\ }\href {\doibase 10.1126/science.1236929}
  {\bibfield  {journal} {\bibinfo  {journal} {Science}\ }\textbf {\bibinfo
  {volume} {341}},\ \bibinfo {pages} {632} (\bibinfo {year}
  {2013})}\BibitemShut {NoStop}%
\bibitem [{Note1()}]{Note1}%
  \BibitemOpen
  \bibinfo {note} {As an entanglement witness or a measure of metrological
  enhancement, $\xi $ should be normalized by twice the Bloch vector length
  rather than the number of particles. However, the Bloch vector length at the
  optimal squeezing time approaches $\protect \mathcal {N}/2$ in the large
  $\protect \mathcal {N}$ limit, and thus its omission does not affect the
  derived scaling exponents.}\BibitemShut {Stop}%
\bibitem [{Note2()}]{Note2}%
  \BibitemOpen
  \bibinfo {note} {These scaling exponents are defined formally by $\nu
  =\protect \qopname \relax m{lim}_{\protect \mathcal {N}\rightarrow \infty }d
  \protect \qopname \relax o{log}\protect \mathaccentV {tilde}07E{\xi
  }^2/d\protect \qopname \relax o{log}\protect \mathcal {N}$, $\mu =\protect
  \qopname \relax m{lim}_{\protect \mathcal {N}\rightarrow \infty }d\protect
  \qopname \relax o{log}\protect \mathaccentV {tilde}07E{t}/d\protect \qopname
  \relax o{log}\protect \mathcal {N}$. These derivatives, evaluated at finite
  $\protect \mathcal {N}$, determine the curves in Fig.~\ref
  {fig:numerical_scaling}. The arbitrary choice of $\xi ^2$ instead of $\xi $
  in the definition of $\nu $ is made so that $\mu $ and $\nu $ are equal at
  $\alpha =0$.}\BibitemShut {Stop}%
\bibitem [{\citenamefont {van~den Worm}\ \emph {et~al.}(2013)\citenamefont
  {van~den Worm}, \citenamefont {Sawyer}, \citenamefont {Bollinger},\ and\
  \citenamefont {Kastner}}]{kastner2012}%
  \BibitemOpen
  \bibfield  {author} {\bibinfo {author} {\bibfnamefont {M.}~\bibnamefont
  {van~den Worm}}, \bibinfo {author} {\bibfnamefont {B.~C.}\ \bibnamefont
  {Sawyer}}, \bibinfo {author} {\bibfnamefont {J.~J.}\ \bibnamefont
  {Bollinger}}, \ and\ \bibinfo {author} {\bibfnamefont {M.}~\bibnamefont
  {Kastner}},\ }\href {http://stacks.iop.org/1367-2630/15/i=8/a=083007}
  {\bibfield  {journal} {\bibinfo  {journal} {New J. Phys.}\ }\textbf {\bibinfo
  {volume} {15}},\ \bibinfo {pages} {083007} (\bibinfo {year}
  {2013})}\BibitemShut {NoStop}%
\bibitem [{SOM()}]{SOM}%
  \BibitemOpen
  \href@noop {} {}\bibinfo {note} {See the supplemental material for
  details}\BibitemShut {NoStop}%
\bibitem [{\citenamefont {Li}\ \emph {et~al.}(2012)\citenamefont {Li},
  \citenamefont {Gong}, \citenamefont {Yin}, \citenamefont {Quan},
  \citenamefont {Yin}, \citenamefont {Zhang}, \citenamefont {Duan},\ and\
  \citenamefont {Zhang}}]{TongcangLi2012}%
  \BibitemOpen
  \bibfield  {author} {\bibinfo {author} {\bibfnamefont {T.}~\bibnamefont
  {Li}}, \bibinfo {author} {\bibfnamefont {Z.-X.}\ \bibnamefont {Gong}},
  \bibinfo {author} {\bibfnamefont {Z.-Q.}\ \bibnamefont {Yin}}, \bibinfo
  {author} {\bibfnamefont {H.~T.}\ \bibnamefont {Quan}}, \bibinfo {author}
  {\bibfnamefont {X.}~\bibnamefont {Yin}}, \bibinfo {author} {\bibfnamefont
  {P.}~\bibnamefont {Zhang}}, \bibinfo {author} {\bibfnamefont {L.-M.}\
  \bibnamefont {Duan}}, \ and\ \bibinfo {author} {\bibfnamefont
  {X.}~\bibnamefont {Zhang}},\ }\href {\doibase 10.1103/PhysRevLett.109.163001}
  {\bibfield  {journal} {\bibinfo  {journal} {Phys. Rev. Lett.}\ }\textbf
  {\bibinfo {volume} {109}},\ \bibinfo {pages} {163001} (\bibinfo {year}
  {2012})}\BibitemShut {NoStop}%
\bibitem [{\citenamefont {Gong}\ and\ \citenamefont {Duan}(2013)}]{Gong13}%
  \BibitemOpen
  \bibfield  {author} {\bibinfo {author} {\bibfnamefont {Z.-X.}\ \bibnamefont
  {Gong}}\ and\ \bibinfo {author} {\bibfnamefont {L.-M.}\ \bibnamefont
  {Duan}},\ }\href {http://stacks.iop.org/1367-2630/15/i=11/a=113051}
  {\bibfield  {journal} {\bibinfo  {journal} {New J. Phys.}\ }\textbf {\bibinfo
  {volume} {15}},\ \bibinfo {pages} {113051} (\bibinfo {year}
  {2013})}\BibitemShut {NoStop}%
\bibitem [{\citenamefont {Neyenhuis}\ \emph {et~al.}(2016)\citenamefont
  {Neyenhuis}, \citenamefont {Smith}, \citenamefont {Lee}, \citenamefont
  {Zhang}, \citenamefont {Richerme}, \citenamefont {Hess}, \citenamefont
  {Gong}, \citenamefont {Gorshkov},\ and\ \citenamefont
  {Monroe}}]{Neyenhuis16}%
  \BibitemOpen
  \bibfield  {author} {\bibinfo {author} {\bibfnamefont {B.}~\bibnamefont
  {Neyenhuis}}, \bibinfo {author} {\bibfnamefont {J.}~\bibnamefont {Smith}},
  \bibinfo {author} {\bibfnamefont {A.~C.}\ \bibnamefont {Lee}}, \bibinfo
  {author} {\bibfnamefont {J.}~\bibnamefont {Zhang}}, \bibinfo {author}
  {\bibfnamefont {P.}~\bibnamefont {Richerme}}, \bibinfo {author}
  {\bibfnamefont {P.~W.}\ \bibnamefont {Hess}}, \bibinfo {author}
  {\bibfnamefont {Z.-X.}\ \bibnamefont {Gong}}, \bibinfo {author}
  {\bibfnamefont {A.~V.}\ \bibnamefont {Gorshkov}}, \ and\ \bibinfo {author}
  {\bibfnamefont {C.}~\bibnamefont {Monroe}},\ }\href
  {https://arxiv.org/abs/1608.00681} {\bibfield  {journal} {\bibinfo  {journal}
  {arXiv:1608.00681}\ } (\bibinfo {year} {2016})}\BibitemShut {NoStop}%
\bibitem [{\citenamefont {Islam}\ \emph {et~al.}(2013)\citenamefont {Islam},
  \citenamefont {Senko}, \citenamefont {Campbell}, \citenamefont {Korenblit},
  \citenamefont {Smith}, \citenamefont {Lee}, \citenamefont {Edwards},
  \citenamefont {Wang}, \citenamefont {Freericks},\ and\ \citenamefont
  {Monroe}}]{islam13}%
  \BibitemOpen
  \bibfield  {author} {\bibinfo {author} {\bibfnamefont {R.}~\bibnamefont
  {Islam}}, \bibinfo {author} {\bibfnamefont {C.}~\bibnamefont {Senko}},
  \bibinfo {author} {\bibfnamefont {W.~C.}\ \bibnamefont {Campbell}}, \bibinfo
  {author} {\bibfnamefont {S.}~\bibnamefont {Korenblit}}, \bibinfo {author}
  {\bibfnamefont {J.}~\bibnamefont {Smith}}, \bibinfo {author} {\bibfnamefont
  {A.}~\bibnamefont {Lee}}, \bibinfo {author} {\bibfnamefont {E.~E.}\
  \bibnamefont {Edwards}}, \bibinfo {author} {\bibfnamefont {C.-C.~J.}\
  \bibnamefont {Wang}}, \bibinfo {author} {\bibfnamefont {J.~K.}\ \bibnamefont
  {Freericks}}, \ and\ \bibinfo {author} {\bibfnamefont {C.}~\bibnamefont
  {Monroe}},\ }\href {\doibase 10.1126/science.1232296} {\bibfield  {journal}
  {\bibinfo  {journal} {Science}\ }\textbf {\bibinfo {volume} {340}},\ \bibinfo
  {pages} {583} (\bibinfo {year} {2013})}\BibitemShut {NoStop}%
\bibitem [{\citenamefont {Braunstein}\ and\ \citenamefont
  {Caves}(1994)}]{PhysRevLett.72.3439}%
  \BibitemOpen
  \bibfield  {author} {\bibinfo {author} {\bibfnamefont {S.~L.}\ \bibnamefont
  {Braunstein}}\ and\ \bibinfo {author} {\bibfnamefont {C.~M.}\ \bibnamefont
  {Caves}},\ }\href {\doibase 10.1103/PhysRevLett.72.3439} {\bibfield
  {journal} {\bibinfo  {journal} {Phys. Rev. Lett.}\ }\textbf {\bibinfo
  {volume} {72}},\ \bibinfo {pages} {3439} (\bibinfo {year}
  {1994})}\BibitemShut {NoStop}%
\bibitem [{\citenamefont {T\'oth}\ and\ \citenamefont
  {Petz}(2013)}]{PhysRevA.87.032324}%
  \BibitemOpen
  \bibfield  {author} {\bibinfo {author} {\bibfnamefont {G.}~\bibnamefont
  {T\'oth}}\ and\ \bibinfo {author} {\bibfnamefont {D.}~\bibnamefont {Petz}},\
  }\href {\doibase 10.1103/PhysRevA.87.032324} {\bibfield  {journal} {\bibinfo
  {journal} {Phys. Rev. A}\ }\textbf {\bibinfo {volume} {87}},\ \bibinfo
  {pages} {032324} (\bibinfo {year} {2013})}\BibitemShut {NoStop}%
\end{thebibliography}
\end{document}